%% file: new.tex
\documentclass[sigconf]{acmart}

\usepackage{xcolor}
\usepackage{graphicx}
\usepackage{adjustbox}
\usepackage{tabularx}
\usepackage{booktabs} 
\usepackage[table]{xcolor}
\usepackage{multirow}
\usepackage{subcaption}
\usepackage{url}
\usepackage{tablefootnote}
\usepackage{threeparttable}
\usepackage{tabularx}  
\usepackage{makecell}
\usepackage[utf8]{inputenc}  
\usepackage{verbatim}

\usepackage{listings}
\usepackage{xcolor}
\usepackage{anyfontsize}
\usepackage{hyperref}
\usepackage{enumitem}
\usepackage{nameref}
\usepackage{array}
\usepackage{xcolor}
\usepackage{makecell}
\usepackage{threeparttable}

\usepackage[ruled,vlined,linesnumbered]{algorithm2e}
\SetKwInput{KwIn}{Input}
\SetKwInput{KwOut}{Output}

\definecolor{dlblue}{RGB}{40,98,180}
\definecolor{dlgreen}{RGB}{20,130,80}
\definecolor{dlorange}{RGB}{180,100,20}
\definecolor{dlpurple}{RGB}{120,70,160}
\definecolor{figsoftbg}{RGB}{246,249,252}

\newcommand{\dlbase}[1]{\texttt{\textcolor{dlblue}{#1}}}
\newcommand{\dlpath}[1]{\texttt{\textcolor{dlgreen}{#1}}}
\newcommand{\dlparam}[1]{\texttt{\textcolor{dlorange}{#1}}}
\newcommand{\dlnested}[1]{\texttt{\textcolor{dlpurple}{#1}}}

\usepackage{hyperref}

\newcommand{\tool}{\textsc{AdHive}\xspace}

\usepackage{xurl}

\usepackage[normalem]{ulem}

\definecolor{lightgray}{gray}{0.9}
\lstdefinestyle{customjavastyle}{
    language=Java,
    basicstyle=\scriptsize\ttfamily,
    keywordstyle=\bfseries,
    stringstyle=\color{red},
    commentstyle=\color{green!50!black},
    numbers=left,
    numberstyle=\tiny\color{gray},
    stepnumber=1,
    numbersep=10pt,
    showspaces=false,
    showstringspaces=false,
    showtabs=false,
    frame=single,
    tabsize=4,
    breaklines=true,
    breakatwhitespace=false,
    backgroundcolor=\color{lightgray},
    keywordstyle = {\color{red}},
    keywordstyle = [2]{\color{blue}},
    keywords = {},
    otherkeywords = {isARU,true},
    morekeywords = [2]{setIsAgeRestrictedUser,setMetaData}
}

\usepackage{pifont}

\newcommand{\ignore}[1]{}

\newcommand{\ws}[1]{\textbf{\textcolor{blue}{#1}}}

\usepackage{listings}
\usepackage{amsmath}       
\usepackage{algpseudocode} 

\newcommand{\fraud}{semi-drive-by ads}
\newcommand{\fraudname}{semi-drive-by}

\newcommand{\para}[1]{\vspace{2pt}\noindent\textbf{#1.~}}

\newenvironment{packeditemize}{
	\begin{list}{$\bullet$}{
			\setlength{\labelwidth}{4pt}
			\setlength{\itemsep}{0pt}
			\setlength{\leftmargin}{\labelwidth}
			\addtolength{\leftmargin}{\labelsep}
			\setlength{\parindent}{0pt}
			\setlength{\listparindent}{\parindent}
			\setlength{\parsep}{0pt}
			\setlength{\topsep}{1pt}}}{\end{list}}

\AtBeginDocument{%
  }
\setcopyright{none}

\begin{document}

\title{When Ad Networks Misbehave: Understanding Risks of \textit{Semi-Drive-By} Splash Ads}

\author{Song Wu}
\affiliation{%
  \institution{Independent Researcher}
  \city{ChongQing}
  \country{China}
  \authornote{Song Wu conducted this work while he was a remote intern in Xueqiang Wang's group.\\ 
  $^\dag$ Yinfeng Cao is the corresponding author.\\
  $^\ddag$ A concise version of this work was published at IEEE S\&P'2026 at \textcolor{teal}{\url{https://sp2026.ieee-security.org/downloads/posters/sp2026posters-final91.pdf}.
  }
  }
}
\email{researchersongwu@gmail.com}

\author{Bo Wang}
\affiliation{%
  \institution{Independent Researcher}
  \city{Hang Zhou}
  \country{China}
}
\email{wan.id.none@gmail.com}

\author{Yifan Zhang}
\affiliation{%
  \institution{San Diego State University}
  \city{San Diego}
  \country{USA}
}
\email{yzhang24@sdsu.edu}

\author{Yinfeng Cao}
\affiliation{%
  \institution{The Hong Kong Polytechnic University}
  \city{Hong Kong}
  \country{China}
}
\email{csyfcao@comp.polyu.edu.hk }
\author{Xueqiang Wang}
\affiliation{%
  \institution{University of Central Florida}
  \city{Orlando}
  \country{USA}
}
\email{xueqiang.wang@ucf.edu}

\begin{abstract}

We investigate the mobile \emph{splash ads} ecosystem, i.e., full-screen advertisements shown at app launch, where monetization relies on interaction signals that are difficult to verify end-to-end. This ad format is especially sensitive because it sits at the boundary between app startup and user navigation, where incidental touches and sensor-driven callbacks are common yet easy to misattribute as engagement. 
Prior work has largely framed mobile ad fraud as a publisher-side problem, while a small number of studies have attributed fraudulent operations to embedded ad libraries. Yet a distinct risk remains underexplored in splash advertising: ad SDKs control both how interaction signals are interpreted and how the resulting events are measured and reported, creating an opportunity to reinterpret ambiguous user or device signals as valid advertising interactions.

We uncover a previously less-known form of fraud at the ad-network layer in which splash ads are triggered not by intentional user actions but by \emph{incidental or indirect interactions}, a behavior we term \emph{semi--drive-by} splash ads. By translating non-ad interactions into billable engagement events, ad networks can systematically inflate performance metrics, overcharge advertisers, and erode user trust while providing little or no real user interest.

To expose this behavior in the wild, we design \tool{}, an automated honeypot-like analysis framework capable of inducing evasive \textit{splash-ad delivery and landing behaviors} under realistic device conditions. Unlike traditional VM-based approaches, \tool{} reproduces human-like activity through LLM-generated usage traces and sensor dynamics, enabling execution paths that remain hidden under conventional analysis environments.  

Our large-scale measurement across thousands of popular Android applications demonstrates that semi-drive-by splash ads are widespread and are often triggered by subtle environmental signals such as minor sensor variations. We further confirm real-world impact by working with one of China’s largest advertisers to identify multiple ad networks engaging in this fraud, which led to enforced repayments of about 4 million Yuan ($\approx$US\$600{,}000).

\end{abstract}


\maketitle

\section{Introduction}
\label{intro}

The mobile \emph{splash ads} ecosystem---full-screen advertisements shown at app launch---has become highly centralized and opaque, while monetization increasingly hinges on interaction signals that are difficult to verify end-to-end. In practice, ad networks simultaneously act as traffic distributors and measurement authorities: they provide SDKs embedded in third-party apps to govern ad delivery, interaction tracking, DeepLink redirection, and billing. This coupling places advertisers, networks, developers, and users into a pipeline where the network’s SDK mediates what constitutes a ``valid'' engagement and how it is reported.

Prior work on mobile ad fraud has largely focused on a publisher-centric threat model, in which the app developer or publisher is the adversary and defenses therefore concentrate on the application layer. Existing studies examine fraudulent traffic generated by bots or click farms~\cite{adsherlock,fcfraud,miller2011whats,xu2014click,2012click,zhang2008detecting}, deceptive ad placement and UI manipulation~\cite{liu2014decaf,DBLP:conf/sigsoft/DongWLGBLXK18}, and programmatically generated or replayed ad interactions~\cite{crussell2014madfraud,madlife,clicktok}.
Other work attributes automated clicks or forced redirections to the in-app modules responsible for generating them~\cite{kim2021abuser}.
Collectively, however, these studies primarily examine the source, authenticity, or presentation of ad interactions rather than how ad-network SDKs themselves interpret incidental or indirect events as valid, billable engagement.
This leaves a critical gap because ad networks occupy a privileged position with unilateral control over (i) the SDK logic that defines engagement, (ii) the telemetry and attribution signals reported upstream, and (iii) the rollout surface (region, device, campaign) used to selectively enable behaviors. In concentrated markets such as China’s mobile splash-ads ecosystem, these providers have both the economic leverage and technical control to deploy deceptive practices that directly distort advertiser spend and measurement.
To our knowledge, however, few studies have systematically examined ad networks themselves as a source of ad fraud, particularly how ad network-controlled SDKs can manipulate or reinterpret user interactions to generate billable engagement.

\para{A new form of threat} In the splash-ads pipeline, we observe a recurring but underexplored pattern: an app-launch ad can open its landing page even when the user never deliberately interacts with the ad. Such activations may be induced by incidental context during normal phone use, including but not limited to small device motions, scrolling, or unrelated touches, yet they are still logged and billed as valid engagement, effectively turning ambient behavior into monetizable signals and enabling engagement inflation that is hard to verify end-to-end.

We term this mechanism \emph{semi-drive-by} splash ads: advertisements whose landing (or click-equivalent engagement) is triggered by \emph{incidental or indirect interactions} rather than deliberate ad-directed input. A prominent instance is \emph{shake-to-trigger} ads, where an overly sensitive SDK threshold misclassifies ordinary movements (walking, handling the phone, riding in a car) as intentional ``shake'' actions and immediately opens the advertiser’s landing page. Because coerced activation and genuine intent have fundamentally different commercial value, this misattribution inflates engagement statistics, wastes ad budget, and corrupts performance analytics while remaining difficult for advertisers to audit.

\para{\textit{Semi-drive-by} fraud is hard to detect} Detecting such frauds at scale is difficult because malicious SDKs are engineered to evade standard analysis pipelines. They often deactivate or mask suspicious logic once virtualization, instrumentation, or debugging artifacts are detected, causing conventional VM-based sandboxes to observe only benign execution. Their trigger conditions may also be stateful and long-horizon, requiring realistic app usage sequences, sustained motion or sensor dynamics, or specific timing patterns before the landing behavior is enabled. Yet the behavior is not always strictly gated: some ads may also be triggered casually during seemingly normal user interaction, which makes the triggering logic harder to model with short scripted tests. In addition, ad networks may activate such behavior selectively by region, campaign, time window, or device profile, making sporadic audits inconsistent and difficult to reproduce. Together, these properties cause both static inspection and naive dynamic testing to systematically miss \textit{semi-drive-by} triggers.

Existing solutions are, unfortunately, ineffective in this setting. Most prior defenses assume a malicious publisher and therefore focus on app-layer indicators such as abnormal click or impression statistics, developer-inserted manipulation, or suspicious UI workflows. Semi-drive-by fraud violates this assumption because the engagement event can be generated inside the network-controlled SDK and then reported upstream as legitimate attribution, leaving limited evidence in developer-authored code and making purely app-centric signals ambiguous. Static analysis is further weakened by dynamic gating and server-side switches that are invisible without execution, while standard dynamic testing is precisely what anti-analysis checks and long-horizon trigger requirements are designed to defeat. As a result, current monitoring pipelines struggle to distinguish genuine user intent from coerced activation and tend to underestimate both prevalence and advertiser-side harm.
We provide a broader comparison with prior publisher-centric ad-fraud and evasive behaviors in Section~\ref{related_work}.

\para{Our approach} We design and implement \tool{}, an automated honeypot framework for dynamically detecting semi-drive-by fraud in splash ads. \tool{} couples an environment-hardened emulator, which is built to withstand common SDK anti-analysis checks, with realistic device profiles, long-horizon usage traces, and motion inputs that are necessary to induce hidden trigger paths. On top of this execution layer, \tool{} provides an LLM-assisted log analysis pipeline that identifies suspicious DeepLink and landing behaviors and attributes them to specific SDK logic with high precision. Using \tool{}, we perform large-scale measurement over popular Android applications in the Chinese market to quantify the prevalence, activation conditions, and ecosystem structure of semi-drive-by behavior.

\para{Measuring \textit{semi-drive-by} splash ads at scale} We crawl non-game apps from four major Android markets, XIAOMI, VIVO, YINGYONGBAO (YYB), and WANDOUJIA (WDJ)\footnote{XIAOMI with \url{https://m.app.mi.com}; VIVO with \url{https://h5.appstore.vivo.com.cn}; YYB with \url{https://sj.qq.com}; WDJ with \url{https://wandoujia.com}.}, obtaining \emph{32{,}758} distinct packages. After filtering apps without network permissions and removing those incompatible at runtime, the final test set contains \emph{31{,}823} apps. We then run 24-hour, uninterrupted launch-phase testing under realistic sensor profiles and usage traces. We identify semi-drive-by behavior in \emph{261}, \emph{128}, \emph{198}, and \emph{191} apps from the VIVO, XIAOMI, WDJ, and YYB markets, respectively, corresponding to prevalences of \emph{4.96\%}, \emph{2.13\%}, \emph{1.70\%}, and \emph{2.15\%}. We further compare matched app cohorts under city-specific network and motion conditions and observe more frequent activation in second-tier cities than in first-tier cities. To validate attribution, security experts manually reverse-engineer a random sample of \emph{100} flagged apps and confirm that all cases originate from SDK logic rather than developer-authored code. Together, these results establish both the prevalence of semi-drive-by splash ads and the causal role of network-controlled SDKs.

\para{Practical impact} We collaborated with a major advertiser, \textit{Company~A}, after it observed large inconsistencies in conversion metrics. Our investigation confirmed systematic misreporting consistent with deceptive interaction triggers. Two leading ad networks acknowledged the behavior after disclosure; one returned 4 million Yuan (USD~600{,}000), which is 18\% of the total fees paid to that network during the period.

\para{Contributions}
This paper makes three contributions:
\begin{packeditemize}
    \item We identify a previously unreported form of ad-network fraud, called \textit{semi-drive-by} splash ads, in which advertisements are triggered by incidental or indirect interactions and billed as legitimate user engagement.
    \item We present \tool{}, an automated and environment-resilient honeypot system that detects \textit{semi-drive-by} ads by combining hardened execution, realistic motion and usage synthesis, and learning-assisted DeepLink and landing-behavior analysis.
    \item We conduct the first large-scale measurement of \fraud{}, quantify market- and region-level prevalence, and analyze the incentives and rollout strategies that sustain the behavior, including real-world remediation following disclosure.
\end{packeditemize}

\para{Responsible disclosure}  Our study follows established community practice for measurement research. We do not collect production user data; instead, we use simulated traces and controlled environments to trigger and observe SDK behaviors. We coordinated responsible disclosure with affected parties, notified implicated networks, received acknowledgements from multiple vendors, and observed concrete remediation, including the refund described above. Further details on safeguards, reporting timelines, and responsible measurement are provided in \nameref{appendix:ethics}.

\section{Technical Warm-ups}
\label{background}

We introduce the role of ad SDKs in mobile advertising and the interaction mechanisms of splash ads.

\begin{figure}[h!]
    \centering
    \includegraphics[width=0.90\linewidth]{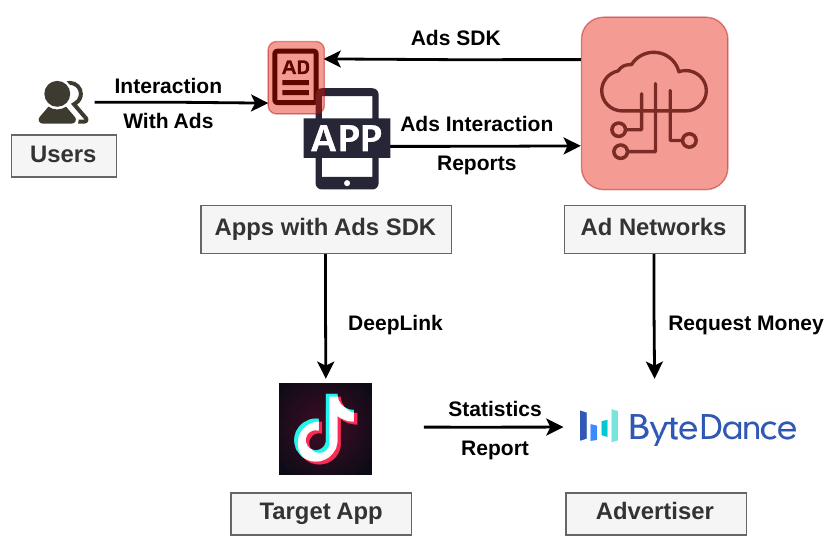}
    \caption{An interaction flow of mobile ads.}
    \label{fig:ecosys1}
\end{figure}

\para{Ad SDKs in mobile advertising}
Ad SDKs are a core component of modern mobile advertising. Large ad platforms provide such SDKs for third-party apps to integrate, enabling app developers to monetize their applications while delegating much of the ad-delivery and measurement logic to the ad network. In this arrangement, developers typically decide where and when an advertisement is shown, whereas the ad network, through its SDK, controls ad selection, interaction capture, redirection, and engagement reporting. Figure~\ref{fig:ecosys1} illustrates this interaction and reporting pipeline among the major stakeholders in the mobile ads ecosystem.

When a user interacts with an advertisement, the SDK embedded in the host app captures the event and takes control of the subsequent flow. It typically redirects the user to the advertiser's landing page or target app via a DeepLink carrying attribution-related metadata, while simultaneously reporting the interaction details to the ad network's backend servers. These records are then used to validate engagement, compute conversion metrics, and support billing between the ad network and the advertiser. Although advertisers may obtain partial feedback from the landing destination or the advertised app, they generally do not directly observe the original user-side interaction. As a result, the ad SDK becomes the critical control point in the measurement pipeline: it mediates what is treated as a valid ad interaction, how that interaction is attributed, and how it is reported upstream.

\para{Splash-Ad interactions}
Splash ads, also known as splash screens or welcome-page ads, are full-screen advertisements shown immediately when an app is launched, before the main interface becomes available. Because they occupy the entire screen during app initialization, they provide immediate exposure and user attention at a particularly sensitive stage of app usage. In a standard workflow, a splash ad should redirect the user to an external landing page, target app, or web page only after an explicit ad-related interaction.

In practice, splash ads expose multiple interaction modes, most commonly tap, swipe, and shake. Tap- and swipe-based interactions rely on direct on-screen input, whereas shake-based interactions depend on device motion sensors such as accelerometers and gyroscopes. This makes shake-to-trigger fundamentally different from ordinary touch-based input: instead of observing an explicit screen action, the system must infer user intent from physical motion. Under benign settings, the detection threshold should therefore be calibrated carefully so that ordinary movements, such as picking up the phone, walking, or riding in a vehicle, are not misclassified as intentional ad engagement.

From a measurement perspective, these interaction modes define the boundary between deliberate user engagement and incidental behavior. Touch-based triggers are more explicit to interpret, whereas sensor-mediated triggers introduce a broader ambiguity between genuine intent and accidental motion. This ambiguity makes splash ads a particularly important setting for studying whether weak or incidental signals can be transformed into externally visible ad redirections and upstream engagement records.

\ignore{
\begin{figure*}[t]
    \centering

    \begin{subfigure}[t]{0.31\linewidth}
        \centering
        \includegraphics[
            height=6 cm,
            width=\linewidth,
            keepaspectratio
        ]{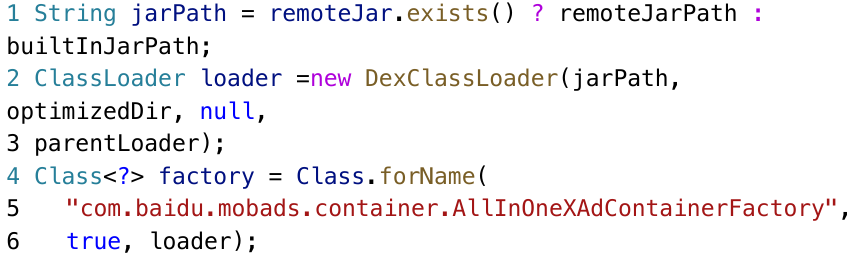}
        \caption{Dynamic JAR loading logic.}
        \label{fig:splashad_dynamic_load}
    \end{subfigure}%
    \hfill
    \begin{subfigure}[t]{0.39\linewidth}
        \centering
        \includegraphics[
            width=\linewidth,
            keepaspectratio
        ]{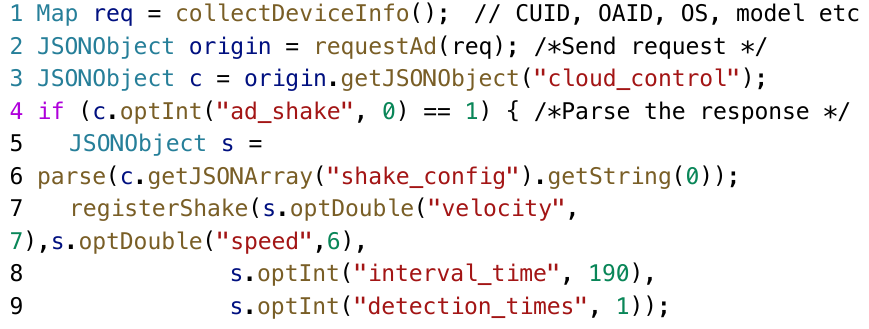}
        \caption{Splash-ad request and response parsing logic.}
        \label{fig:splashad_load}
    \end{subfigure}%
    \hfill
    \begin{subfigure}[t]{0.24\linewidth}
        \centering
        \includegraphics[
            width=\linewidth,
            keepaspectratio
        ]{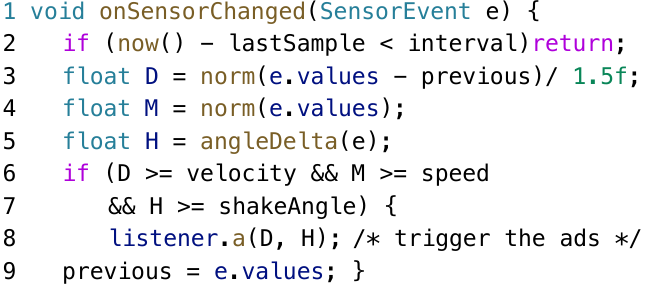}
        \caption{Sensor-based motion detection logic.}
        \label{fig:dynamic_threshold}
    \end{subfigure}

    \caption{Core logic of sensor-triggered splash ads:
    (a) dynamic loading of the advertising component,
    (b) splash-ad request construction and response parsing, and
    (c) evaluation of sensor events against dynamically configured
    triggering conditions.}
    \label{fig:splash_ad_logic}
\end{figure*}
}

\section{Motivation and Challenges}
\label{sec:motivation}

\begin{figure}[ht]
\centering

\begin{subfigure}[b]{0.46\textwidth}
    \begin{lstlisting}[style=customjavastyle]
ClassLoader loader = new DexClassLoader(jarPath, ...);
Class<?> factory = Class.forName("com.baidu.mobads.container.AllInOneXAdContainerFactory", true, loader);
    \end{lstlisting}
    \vspace{-4pt}
    \caption{Dynamic loading of ads container.}
    \label{fig:splashad_dynamic_load}
\end{subfigure}

\begin{subfigure}[b]{0.46\textwidth}
    \begin{lstlisting}[style=customjavastyle,firstnumber=3]
Map req = collectDeviceInfo(); //collect hardware, kernel_qemu, OS, model, etc.
JSONObject o = requestSplashAd(req);
JSONObject c = o.getJSONObject("cloud_control");
if (c.optInt("ad_shake", 0) == 1) {
    JSONObject s = parse(c.getJSONArray("shake_config").getString(0));
    registerShake(s.optDouble("velocity", 7), s.optDouble("speed", 6));
}
    \end{lstlisting}
    \vspace{-4pt}
    \caption{Splash-ad request and response parsing logic.}
    \label{fig:splashad_load}
\end{subfigure}

\begin{subfigure}[b]{0.46\textwidth}
    \begin{lstlisting}[style=customjavastyle,firstnumber=10] 
void onSensorChanged(SensorEvent e) {
    float D = norm(e.values - previous) / 1.5f;
    float M = norm(e.values);
    if (D >= velocity && M >= speed) {
        listener.a(D, H); // trigger ads
        previous = e.values;
    }
}
    \end{lstlisting}
    \vspace{-4pt}
    \caption{Triggering of splash ads based on sensor changes.}
    \label{fig:dynamic_threshold}
\end{subfigure}
\caption{Code snippets related to shake-to-trigger splash ads.}
\label{fig:code-example}
\end{figure}

\subsection{Motivating Example}

\textit{WiFi Master Key} (version 5.1.68, package \texttt{com.snda.wifilocating}) is a popular Wi-Fi sharing and connection app, with monthly active users reported to exceed 520 million~\cite{chinadaily_wifimasterkey_2016}.
We found that this app, for monetization, displays shake-to-trigger splash ads.
When tested on a physical phone, such ads are triggered automatically and redirect users to external landing pages or target apps that the ads aim to promote, even when users do not issue any explicit shake events or when the phone is placed stationary on a table.
By contrast, when the app is executed in an Android emulator, such ads are not displayed to users.
Our review of the app code found that it integrates the Baidu MobAds SDK for serving splash ads.
Specifically, as shown in Figure~\ref{fig:code-example}, the SDK first dynamically loads an ad container responsible for requesting, parsing, and displaying ads (Lines 1--2).
Then, the SDK collects a variety of device and runtime information, including information often used to identify emulated environments (e.g., \texttt{hardware}, \texttt{kernel\_qemu}, \texttt{OS}, and \texttt{model}), and uses this information to request splash ads (Lines 3--4).
Upon receiving the ad content, the SDK parses it to check whether it carries parameters for cloud control and whether it is a shake-to-trigger ad (Lines 5--6).
If so, the SDK extracts the shake configuration (\texttt{shake\_config}) and registers an ad-triggering event by specifying the velocity and speed thresholds for triggering the ad (Lines 7--8).
Note that the velocity and speed values either come from the ad response (i.e., are determined by the ad network through cloud control) or are set to local defaults of 7.0 and 6.0, respectively.
When the splash ad is displayed, the SDK relies on a callback that listens for sensor changes to determine whether the ad should be triggered (Line 10).
Upon changes in sensor values, the \texttt{onSensorChanged} function is invoked, which extracts velocity and speed values from the current sensor readings.
If these values exceed the thresholds specified by the ad content, the splash ad is triggered (Lines 13--16).
Our analysis of the app shows that the thresholds (setup in Line 8) are substantially lower than the recommended thresholds in advertising standards such as TC260~\cite{TC2602025ShakeAds}, leading to overly sensitive triggers, which explains why splash ads are automatically triggered without explicit user interaction (e.g., when users walk while holding their phones without explicitly shaking them).
We call such splash ads \textit{semi-drive-by splash ads}.

\para{Characterizing semi-drive-by ads}
\label{Characterizing}
The above motivating example helps us to better characterize semi-drive-by ads -- an essential first step toward detecting them at scale.
(1) \textit{Dynamic code loading.}
In the motivating example, the SDK dynamically loads an ad container and receives trigger parameters through \texttt{shake\_config}.
Without inspecting the app's runtime behavior, such semi-drive-by ads are difficult to capture.
(2) \textit{Overly sensitive trigger logic.}
The Baidu MobAds SDK reads motion sensor data and uses default overly low thresholds, causing unintentional device movements to be treated as shake-to-trigger interactions. 
Although such movements do not reflect clear user intent, they are still reported by the SDK as valid engagement and contribute to advertiser billing.
(3) \textit{Evasive behavior across runtime environments and users.}
The SDK collects a range of device information commonly used to detect virtualized environments, and our observations show that such information is later used to determine whether ads should be delivered to the user -- a type of evasive behavior that has been observed across a variety of malicious apps, such as malware.
Such evasive behavior may extend beyond virtualized environments to other differentiating factors of users, such as geographic information, user usages.

\subsection{Challenges to Detect At Scale}
\label{subsec:challenges}

Bringing to light an in-depth understanding of the prevalence of semi-drive-by ads requires scalable analysis that addresses the above characteristics of such ads, posing several challenges to straightforward app analysis.

\para{C1: Large-scale analysis cannot rely on manual or device-heavy testing}
Since deceptive ad behaviors may be loaded or updated at runtime, static analysis alone is insufficient and dynamic analysis becomes necessary. However, scaling dynamic analysis with large numbers of physical devices is expensive and difficult to maintain. A practical solution must therefore support large-scale testing without relying on costly device-heavy deployment.

\para{C2: VM-based testing misses evasive \fraud{}}\label{para:c2-vm-evasion}
A natural way to scale dynamic analysis is to use virtual machines or emulators, as in prior work~\cite{sahin2018proteus, Li2024AndroidCatMouse, specter2025fingerprinting}. According to our preliminary analysis, however, \fraud{} are highly evasive: they may disappear in virtual environments, on freshly initialized devices, or under certain regional conditions.  Consequently, merely scaling up dynamic testing in a VM is insufficient. The analysis environment must
remain realistic across the execution stack, from Android framework APIs to native and kernel-exposed signals, so that evasive SDK logic is not suppressed before \fraud{} can be observed. Moreover, because such evasion may also depend on user usage, the analysis environment should preserve realistic usage traces rather than resemble a freshly initialized device.

\para{C3: Semi-drive-by behaviors are mixed with normal app behavior}
Even when deceptive behavior is triggered, identifying it is non-trivial. As illustrated in Figure~\ref{fig:ecosys1}, ad SDKs reuse mechanisms that are also common in benign app workflows, such as DeepLink and inter-app jumps. Simply observing an external redirection is therefore insufficient to conclude that \fraud{} have occurred. An effective analysis framework must distinguish ad-related triggering, redirection, and reporting behaviors from the large volume of normal app and device activity.

\section{Our Detection Tool: \tool{}}
\label{system}

\begin{figure*}[t]
    \centering
    \includegraphics[
        width=0.8\textwidth,
        keepaspectratio,
        alt={}
    ]{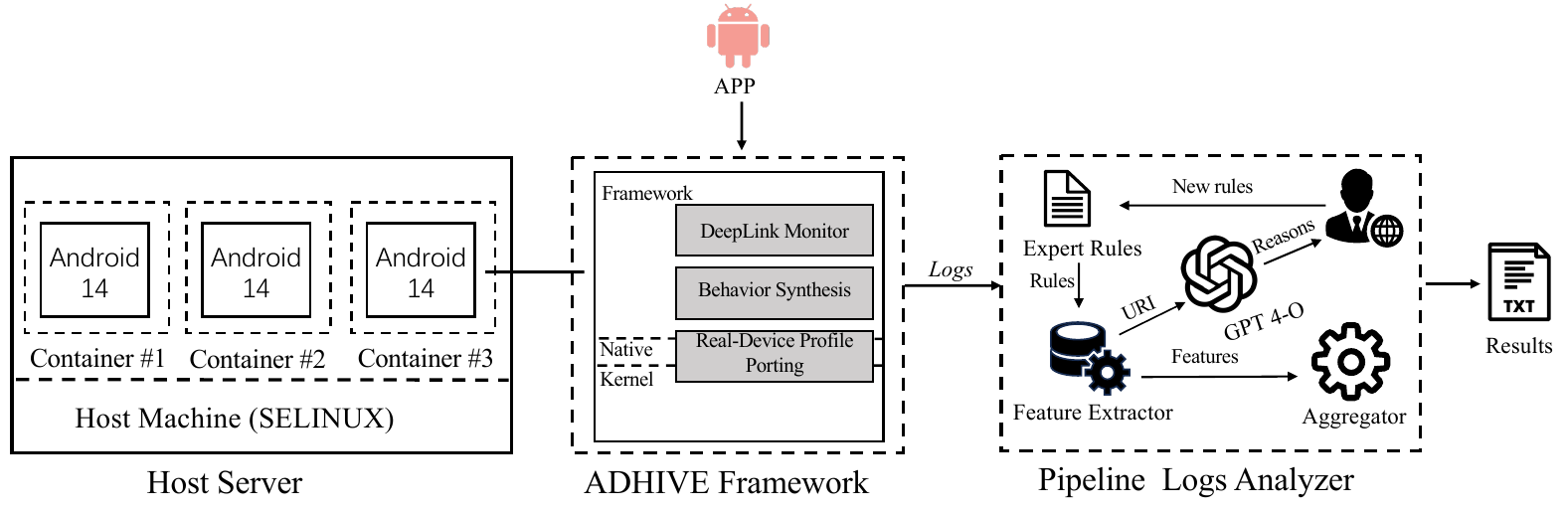}
    \caption{Architecture of AdHive.}
    \label{fig:adhivsystem}
\end{figure*}

We propose \tool{} (Figure~\ref{fig:adhivsystem}), an automated framework for measuring and detecting \fraud{} on Android. The main challenge is that such behaviors are conditionally exposed: ad SDKs may hide suspicious logic under unrealistic environments, require plausible user traces before activating monetized paths, and trigger ads through ordinary-looking DeepLink flows.

To address this, \tool{} builds a realistic honeypot runtime. It ports coherent real-device profiles into virtualized Android, generates believable usage traces and controllable sensor inputs, and analyzes logs to separate normal app behavior from ad-triggered transitions. The log engine jointly examines DeepLink structures, attribution parameters, and timing context.

The key idea is to induce \fraud{} under realistic conditions while instrumenting system-level choke points that every DeepLink invocation must pass through. Since we do not perform deliberate ad-directed interactions and keep injected motion below conservative engagement thresholds, ad-related events observed at these points provide strong evidence of SDK-side automatic triggering rather than genuine user intent.

\subsection{Design Overview}
\label{sec:designoverview}

\para{System components} \tool{} consists of three tightly coupled components. The first is a \emph{real-device profile porting module}, which reconstructs a coherent execution environment by replaying cross-layer signals extracted from a physical handset into a containerized Android runtime and aligning the surrounding runtime state to remain consistent with that profile. The second is a \emph{behavior synthesis module}, which creates realistic usage traces and controllable sensor inputs so that the runtime appears actively used rather than freshly instantiated. The third is an \emph{ad-related event attribution module}, involving a DeepLink monitor and a pipeline log analyzer, which processes execution logs, DeepLink intents, and transition metadata to determine whether an observed behavior is normal or ad-related. These components play different but complementary roles: the first reduces environment-level detectability, the second enables the controlled exposure of behaviorally gated logic, and the third attributes observed transitions to ad-related delivery paths.

\para{Workflow} \tool{} follows a simple measurement pipeline. It first instantiates a containerized Android runtime and ports into it a real-device profile extracted from a physical handset. It then populates the environment with synthetic yet realistic traces of user activity, including APK installation records, contacts, routine interactions, and motion signals for shake-based flows. The target app is executed under this environment while \tool{} monitors DeepLink intent dispatch at a carefully chosen instrumentation point. This design is sufficient because the dispatch point already exposes the key information needed for later attribution, including the DeepLink itself, routing targets, transition context, and ad-related parameters. Finally, the collected events are analyzed to classify each observed transition as normal or ad-related and to extract structured targets and attribution fields. In this way, \tool{} serves not only as a trigger engine for surfacing hidden behaviors, but also as an attribution pipeline for interpreting how those behaviors are delivered.

\para{Threat model} We consider Android apps that integrate third-party advertising SDKs or ad mediation components. The adversary is the ad-side logic embedded in these SDKs, rather than the host app developer alone. Such logic can observe cross-layer execution signals, including device-identity cues, runtime artifacts, and behavioral context, and can conditionally trigger monetization actions only when the surrounding environment appears authentic. Our target behaviors include covert redirections, unintended ad landings, market jumps, mini-program launches, and monetization-driven transitions not clearly attributable to deliberate user engagement. 

In particular, we consider two fraud mechanisms: \textit{direct launches} that occur without any meaningful sensor-driven trigger, and \textit{sensor-mediated launches} that rely on overly sensitive thresholds so that ordinary motion can be misinterpreted as intentional interaction. 

For simplicity, we assume that under a given SDK configuration these two mechanisms are mutually exclusive, since an SDK that already performs direct triggering has little need to additionally hide behind low-threshold sensor logic. We assume the analyst can install and run the target app inside a controlled Android environment, collect the runtime events needed for attribution, and inject user-activity signals such as routine interactions and motion traces. We do not assume access to SDK source code, ad platform backends, or private billing records. Our objective is therefore behavioral: to expose and attribute suspicious ad-related transitions that manifest on-device under realistic conditions.

\subsection{Real-Device Profile Porting}
\label{sec:profile}

This component reconstructs a realistic Android execution substrate by porting a coherent real-device profile from a physical handset into a containerized runtime, with the goal of reducing environment-level detectability. As noted earlier in \S\ref{subsec:challenges}-\textbf{C2}, \fraud{} often disappear in straightforward virtualized environments, suggesting that semi-drive-by logic is protected by environment-gating checks. This observation is consistent with prior work on emulator detection, which shows that apps and SDKs can distinguish emulators from physical devices by probing discrepancies in observable signals~\cite{Jing2014Morpheus}. Our reverse engineering further indicates that ad SDKs inspect not only framework-level artifacts, but also lower-level execution signals exposed through native interfaces and system calls. A believable honeypot therefore requires not isolated spoofing, but a cross-layer profile whose internal signals remain mutually consistent. 

Existing solutions are insufficient for this purpose. 
For instance, CamoDroid~\cite{camodroid} mainly performs instrumentation at the framework layer, which is too narrow to satisfy our cross-layer consistency requirement. VPBOX~\cite{song2021vpbox} moves closer to multi-layer environment simulation, but its signal design is driven primarily by the detection logic of malicious apps rather than the behaviors of ad SDKs, making it less comprehensive for our setting. Moreover, neither system incorporates realistic usage context, which is important because evasive ad SDKs may condition their behaviors not only on device fingerprints but also on whether the runtime appears naturally used.

\para{Core design} Rather than synthesizing fingerprints field by field, \tool{} replays a \emph{real-device profile} into a Redroid-based Android container~\cite{redroid}. To identify which signals should be faithfully reconstructed, we asked three security experts to manually reverse-engineer 30 semi-drive-by samples and analyze the environment-inspection logic used by the embedded ad SDKs. Based on this analysis, we developed a profile generation tool, \textsc{DeviceFP}, which extracts and assembles the signals required to emulate a coherent real-device profile. The profile is a structured bundle of signals extracted from a physical handset, including \texttt{ro.product.*} properties, selected filesystem and shell-command outputs, kernel-visible attributes, returns from Android system services, and the SELinux posture. By treating these signals as a coherent profile rather than a flat list of spoofed values, \tool{} preserves the natural correlations that commercial SDKs may validate during runtime checks.

This design reflects how authenticity checks are typically performed in practice. Ad SDKs and anti-fraud logic often combine multiple weak indicators instead of depending on a single flag. For example, they may compare product properties against service-reported hardware information, inspect filesystem layouts for virtualization artifacts, or cross-check security settings with the claimed device model. Piecemeal spoofing may bypass some individual checks, but it often breaks these cross-layer relations and therefore remains detectable. Profile-level replay addresses this problem by restoring consistency across the execution stack rather than merely changing a few exposed identifiers.

\para{Practical instantiation} In our implementation, the containerized runtime is based on Redroid with Android~14~r45. We use a VIVO S17e running Android~14 (build PD2285B\_A\_141.1.170) as the source device, motivated by its representativeness in the mainstream Android ecosystem and the practical feasibility of firmware extraction and profile replay among major brands.

\subsection{Behavior Synthesis}
\label{subsec-usagetrace}

If real-device profile porting makes the runtime look authentic, this component makes it look \emph{used}. This design is motivated by both prior work and our own observations. Prior work shows that app-usage context is a common fingerprinting surface: 38.46\% of likely fingerprinting SDKs collect app-usage information such as installed apps, foreground apps, or usage statistics~\cite{specter2025fingerprinting}. In our manual analysis, we further observed that a major ad SDK queried APK installation records and that several samples requested or collected contact-related data unrelated to their advertised functionality. These observations suggest that deceptive ad logic may rely not only on static device fingerprints, but also on behavioral context, such as whether the device contains plausible communication records, routine activity traces, and motion patterns consistent with everyday handling. A sterile container may therefore pass some environment checks while still failing to expose behaviorally gated ad logic. 

To address this gap, \tool{} actively populates the runtime with realistic user traces and replayable sensor inputs.

\para{Usage trace synthesis} \tool{} creates synthetic but realistic records of ordinary phone activity, including APK installation records, contacts, and routine interaction traces. Rather than relying on hand-written templates, which quickly become repetitive and statistically brittle, we use an LLM-driven synthesizer to generate diverse and contextually coherent content. The synthesis is constrained by analyst-defined rules such as operator prefixes, business hours, diurnal activity windows, and redaction policies, so that the resulting traces remain plausible without introducing unnecessary privacy risks. The purpose is not to mimic any specific user, but to create a device state that appears naturally lived-in and therefore less likely to suppress fraud-relevant execution paths.

\para{Sensor-driven triggering} Beyond passive traces, some ad flows must be actively exercised through motion-sensitive inputs. This is important for splash ads and shake-triggered interactions, where suspicious behaviors may only appear after receiving accelerometer or rotation signals within a plausible physical range. \tool{} generates continuous and tunable motion traces that remain realistic to the SDK while still enabling controlled triggering.

In our setting, these traces are conservatively bounded within physically plausible motion ranges\footnote{thresholds such as $\leq 15~\mathrm{m/s^2}$ acceleration or $\leq 35^\circ$ rotation,  based on the principle published by TC260 \cite{TC2602025ShakeAds}.}, allowing the system to probe borderline ad-trigger conditions without introducing obviously exaggerated user actions. This makes the resulting observations easier to interpret, since an exposed transition can be attributed to SDK-side trigger logic rather than to an artificially strong stimulus.

\subsection{Ad-Related Event Attribution}
\label{subsec-analyzelog}

Making suspicious behavior appear is only part of the problem; the system must also decide whether an observed transition is actually ad-related. Ordinary app control flow and monetization-driven traffic often share the same underlying mechanisms, especially DeepLinks and intent-based activity launches. A login redirect, an in-app navigation event, and an ad-triggered jump may therefore look similar at the raw log level.

\para{Component inputs} This component analyzes the execution evidence collected during runtime from the DeepLink monitor, including activity launches, intent dispatches, DeepLink URIs, extras, and transition metadata. It attributes each observed transition as either benign app logic or ad-related delivery behavior.

\para{Attribution logic} \tool{} performs attribution via a conservative rule-based analyzer over the collected runtime events. Manual comparisons show that ad-related transitions differ from ordinary app flows along stable dimensions such as scheme ownership, parameter vocabulary, payload length, external-launch behavior, and redirection structure, as summarized in Appendix Figure~\ref{tab:specialdeeplinks}. Based on these observations, the analyzer combines event normalization with an expert rulebook that captures high-confidence signals such as known ad-related schemes, parameter aliases, redirection markers, and normalization rules for noisy attribution fields. GPT-4o is used only off the critical decision path to suggest previously unseen attribution-parameter variants for manual rulebook expansion; final labels are assigned solely by deterministic aggregation rules.

For each event, the analyzer aggregates the matched evidence into a high-confidence attribution result, together with a structured parse of the target destination, routing path, and attribution fields. These attributed traces expose recurring delivery patterns, including super-app launches through \texttt{hap://} links, market transit via app-market or push-service handlers, direct market-download jumps, keyword-based market searches, and WeChat mini-program redirections identifiable through characteristic \texttt{APPID}- and path-based routing.

\subsection{System Integration}
\label{sec:system-discussion}

The three components make \tool{} a measurement system rather than a collection of isolated heuristics. Real-device profile porting establishes a runtime that is difficult to dismiss through environment checks alone. Behavior synthesis and sensor replay then move this runtime from a merely device-consistent state to one that also appears actively used, allowing behaviorally gated ad logic to surface under plausible conditions. The attribution component interprets the resulting transitions and determines whether they correspond to ordinary app behavior or monetization-driven redirection. 

The key point is that these components are sequentially dependent: without a realistic substrate, suspicious logic may remain suppressed; without behavioral stimulation, it may never be triggered; and without attribution, the observed events would remain uninterpreted runtime artifacts. This end-to-end dependency is what distinguishes \tool{} from generic emulator-based testing or log collection. Rather than simply executing apps in a virtual environment, \tool{} reconstructs the conditions under which evasive ad-side fraud becomes observable and then turns the resulting behavior into structured evidence of \fraud{}.

\section{\tool{}: Implementation and Validation}
\label{sec-implem}

\subsection{Implementation}
\label{method_implementation}
As discussed in \S\ref{sec:profile}, exposing environment-dependent semi-drive-by behavior requires shrinking the observable gap between a virtualized analysis runtime and a real user device. We therefore first conducted a manual study of 30 confirmed semi-drive-by samples. Three security experts inspected SDK code paths and runtime traces to identify the environment signals exercised by these SDKs. This analysis revealed six categories of SDK-observable signals: 5 unique identifiers, 358 system properties, 37 device-status signals, 15 file-information signals, 150 application-service signals, and 29 hardware-information signals. We also summarized the corresponding cross-layer invocations in Appendix Table~\ref{tab:file_calls}. Based on this analysis, we built \textsc{DeviceFP}, a tool that extracts replayable real-device profiles.

Based on this analysis and prior work~\cite{specter2025fingerprinting,kondracki2022droid}, we customize Redroid (Android~14~r45) to replay a ported profile across four planes. First, we instantiate \emph{system properties} by creating device-specific \texttt{prop} entries following Google’s official workflow~\cite{androidAddNewDevice}. Second, at the kernel boundary, we intercept key syscalls used to probe files and storage, including \texttt{faccessat}, \texttt{newfstatat}, \texttt{statfs}, and \texttt{openat}. For existence and attribute checks, we either deny VM-revealing paths or substitute attributes recorded from the real device. For reads, we redirect accesses to sanitized snapshots extracted from the physical phone. We also forge command outputs that would expose virtualization and replay non-file identifiers when needed. Algorithm~\ref{alg:file-redirect} summarizes this procedure. Third, inside \emph{Android system services}, we follow the replacement-replay strategy of VPBOX~\cite{song2021vpbox} and replace the return values of services such as Telephony, Build, Sensor, and Package so that API results remain profile-consistent. Finally, to balance realism and compatibility, we keep Android SELinux in \emph{Permissive} mode inside the container, disable host SELinux, and selectively constrain accesses through kernel intercepts. As the source device, we use a VIVO S17e running Android~14 (PD2285B\_A\_141.1.170), as shown in Table~\ref{tab:example}. Additional implementation details and supporting appendix tables are deferred to Appendix~\ref{appendix:impl_supp}.
\begin{algorithm}[t]
\caption{Kernel Function File Redirect}
\label{alg:file-redirect}
\KwIn{$P$: path parameter of this function\\$C$: nonexistent path set in real devices\\$S$: VM detection path set\\$D$: path replacement map}

\KwOut{return value of the modified kernel function}

\ForEach{$subpath \in C$}{
    \If{\textsc{SubIndex}$(P, subpath)$}{
        \Return FILE\_NOT\_FOUND\;
    }
}

\ForEach{$subpath \in S$}{
    \If{\textsc{SubIndex}$(P, subpath)$}{
        $P \gets \textsc{SubPathReplace}(P, subpath,$
        $D[subpath])$\;
        \textbf{break}\;
    }
}

\Return \textsc{OriginalKernelFunction}$(P)$\;
\end{algorithm}

\begin{table*}[t]
  \centering
  \footnotesize
  \caption{Comparison of Emulator Detection Coverage Across Different Techniques}
  \label{tab:detection_coverage}
  \begin{tabular}{c|cccccc|cc}
    \toprule
    \textbf{Type} & \multicolumn{6}{c|}{\textbf{Emulators}} & \multicolumn{2}{c}{\textbf{Tools}} \\
    \cmidrule(lr){1-1} \cmidrule(lr){2-7} \cmidrule(lr){8-9}
    \textbf{Name} & genymotion\_x86 & Redroid\_arm & emulator\_x86 & genymotion\_arm & emulator\_arm & mumu\_arm & VPBOX & \tool{} \\
    \midrule
    Network\cite{Vidas2014SandboxDetection} & 2/5 & 1/5 & 2/5 & 2/5 & 2/5 & 1/5 & 0/5 & 0/5 \\
    Performance\cite{Vidas2014SandboxDetection} & 0/2 & 0/2 & 0/2 & 0/2 & 0/2 & 0/2 & 0/2 & 0/2 \\
    Hardware Components\cite{Vidas2014SandboxDetection} & 2/9 & 4/9 & 1/9 & 2/9 & 1/9 & 0/9 & 2/9 & 0/9 \\
    Hypervisor Heuristic\cite{petsas2014rage} & 2/2 & 0/2 & 2/2 & 1/2 & 1/2 & 0/2 & 0/2 & 0/2 \\
    Instruction-level\cite{shi2019jekyll} & 1/6 & 0/6 & 0/6 & 0/6 & 0/6 & 0/6 & 3/6 & 0/6 \\
    Sensor Event\cite{bordoni2017mirage} & 1/2 & 2/2 & 1/2 & 1/2 & 1/2 & 1/2 & 0/2 & 0/2 \\
    API\cite{Jing2014Morpheus} & 10/47 & 19/47 & 12/47 & 7/47 & 12/47 & 2/47 & 7/47 & 0/47 \\
    System Property
    \cite{Jing2014Morpheus,Vidas2014SandboxDetection}& 4/8 & 1/8 & 4/8 & 2/8 & 4/8 & 0/8 & 0/8 & 0/8 \\
    Dual-instance/Plugin\cite{shi2019jekyll,zheng2018plugin} & 0/4 & 0/4 & 0/4 & 0/4 & 0/4 & 0/4 & 0/4 & 0/4 \\
    SafetyNet\cite{android_safetynet_attestation} & $\times$ & $\times$ & $\times$ & $\times$ & $\times$ & $\times$ & $\times$ & $\checkmark$ \\
    \bottomrule
  \end{tabular}
  \begin{minipage}{\linewidth}
\footnotesize
\textit{Note.} For the results like “X/Y”, Y is the total number of detection heuris\-tics, and X is the number of effective ones. For the results of SafetyNet, $\times$ means a tool successfully detects this virtual environment, and $\checkmark$ means it treats this virtual environment as a genuine Android device.
\end{minipage}
\end{table*}

We implement realistic usage traces by writing directly to Android content providers with appropriate permissions. Contact-related data are injected through \texttt{CallLog.Calls.CONTENT\_URI} and the Contacts provider. We use an LLM to produce seed lists for contact synthesis, including common mainland-China mobile and service-number prefixes as well as Chinese name components. We then append random digits under valid length constraints and randomly combine surnames, compound surnames, and given-name candidates to create synthetic contact entries. Implementation details are provided in our open-science artifacts.

For app-installation records and routine usage traces, we install and periodically run the top 15 apps drawn from QuestMobile's~\cite{QuestMobile2025} 2024 ``TOP50'' leader list, so that the resulting device state is consistent with a naturally used handset rather than a freshly initialized sandbox.

To ensure ad paths appear, we automatically pass first-run privacy dialogs by OCR-matching agreement controls (keywords in the appendix) and grant a curated set of 11 core permissions informed by prior work~\cite{alkinoon2025comprehensive,meng2025assessingprivacycomplianceandroid,rodriguez2025privacy}. 

Because the containerized environment lacks hardware sensors, we emulate sensors at the framework boundary following Android’s sensor model~\cite{androidSensorsOverview} and AOSP Android~14 references~\cite{androidFrameworksBaseR45}. On \texttt{SensorManager} registration we attach a per-listener task that periodically dispatches synthetic \texttt{onSensorChanged} events; on unregistration we tear down the task. The data generator produces continuous, physically plausible time series for Accelerometer, Linear Accelerometer, Rotation Vector, and Gyroscope that keep derived acceleration and rotation below the national thresholds (e.g., $\le 15~\mathrm{m/s^2}$ and $\le 35^\circ$), while smoothing/gravity separation is handled internally.

\begin{table*}[!]
  \centering
  \caption{Effect of Usage Trace Forgery on Fraud Trigger Exposures}
  \label{tab:eval_summary}

    \centering
    \footnotesize
      \centering
      \resizebox{0.8\linewidth}{!}{%
        \begin{tabular}{c|cccccccccccccccccc}
          \toprule
          Launches$->$ & \textbf{5} & \textbf{10} & \textbf{15} & \textbf{20} & \textbf{25} & \textbf{30} & \textbf{35} & \textbf{40} & \textbf{45} & \textbf{50} & \textbf{55} & \textbf{60} & \textbf{65} & \textbf{70} & \textbf{75} & \textbf{80} & \textbf{85} & \textbf{90} \\
          \midrule
          Without Usage Traces & 2 & 2 & 2 & 5 & 7 & 10 & 11 & 13 & 13 & 13 & 14 & 17 & 18 & 18 & 18 & 18 & 18 & 18 \\
          With Usage Traces & 7 & 12 & 16 & 23 & 29 & 37 & 42 & 51 & 62 & 68 & 78 & 80 & 82 & 91 & 96 & 96 & 96 & 96 \\
          \bottomrule
        \end{tabular}%
      }
  \hfill

\end{table*}

Semi-drive-by ads ultimately surface as DeepLink launches, so we instrument the point where these launches become explicit. On Android~14\_r45, we hook the \texttt{executeRequest} method. It is reached in \texttt{ActivityStarter} after target resolution passes through Android's task and package management logic. This hook exposes the caller, the intent parameters, and the URI payload in one place, which greatly simplifies large-scale logging and later analysis.

We implement a three-stage rule-based pipeline, with GPT-4o used only as an auxiliary semantic checker for rule expansion.

\emph{(1) Preprocess \& normalize.} The analyzer queries intent logs from the database and constructs a canonical record (8-item tuple) per event: \{\texttt{appPkg}, \texttt{callingPkg}, \texttt{action}, \texttt{scheme}, \texttt{host}, \texttt{path}, \texttt{uri}, \texttt{ts}\}. Details for raw captured logs and the tuple are provided in Appendix~\ref{appendix:raw_log}. The analyzer then percent-decodes payloads, flattens nested links (e.g., \texttt{tbopen://...h5Url=https://...}), resolves known wrappers such as \texttt{weixin://} and \texttt{tbopen://}, and deduplicates near-identical events.

\emph{(2) Rule-based feature extraction.} We apply expert rules to the tuple described above. These rules capture signals such as scheme ownership mismatch with \texttt{appPkg}; the presence of attribution lexemes (\texttt{channel\_id}, \texttt{ad\_slot}, \texttt{utm\_*}, \texttt{gdt}, \texttt{oaid}, \texttt{click\_ts}); and known ad-distribution schemes (Appendix Figure~\ref{tab:specialdeeplinks}). Pure expert rules may miss non-standard expressions, so GPT-4o is used only as an auxiliary semantic checker to suggest candidate rule expansions, which are manually reviewed before inclusion. More details are provided in Appendix~\ref{tagrule}.

\emph{(3) Rule-based evidence aggregation.} We aggregate the extracted features using deterministic evidence rules; further details are provided in Appendix~\ref{evidence}. The aggregation strategy is intentionally conservative: we assign high-confidence advertisement labels only when multiple pieces of evidence jointly support the classification. We then construct a directed graph from caller to target package, which is used downstream for advertisement attribution and fraud analysis.

        
        
        

\subsection{Validation}
\label{subsec-validation}

\subsubsection{\underline{VM detection evaluation}}
To assess the capability of \tool{} to evade virtual machine (VM) detection, we compare it with VPBOX~\cite{song2021vpbox} and several mainstream emulators, including the Android Emulator~\cite{android_emulator}, ReDroid~\cite{redroid}, Genymotion~\cite{GenymotionSite2025}, and MuMu~\cite{MuMuPlayerSite2025}.
We do not include CamoDroid~\cite{camodroid} because it is a Frida-based script that relies on an emulator, Frida, and root privileges, all of which may introduce detectable artifacts.
The detailed setup of the experimental platforms and configurations is provided in Table~\ref{tab:platform}.

To build a set of testing apps that detect VMs, we ran the finance apps from our crawled dataset (detailed in Section~\ref{subsec:data_collection}) in the LDPlayer emulator~\cite{LDPlayer2026}, resulting in 205 apps that display explicit VM-detection warnings.
Then, we randomly sampled 100 apps from this set to evaluate \tool{}'s capability to evade VM detection, in comparison with the other emulators.
We chose LDPlayer because it is different from all the emulators included in the comparison, which helps reduce potential bias and overlap.
We chose finance apps because of their well-known capability to detect device anomalies (including VMs)~\cite{chen2020empirical,kim2021detecting} in order to protect highly sensitive financial assets.
Each app was executed manually in the above emulators and \tool{}, and a run was considered normal if the app launched successfully (e.g., no crashes and the UI rendered correctly) and no VM-detection warnings appeared (i.e., the emulator successfully evaded VM detection).
We report the results along three axes.

%
%

\begin{table*}[!htbp]
  \centering
  \footnotesize
  \caption{Abnormal Rate of Different Emulators and Tools}
  \label{tab:emulator_abnormal_rate}
  \begin{tabular}{c|cccccc|cc}
    \toprule
    \textbf{Type} & \multicolumn{6}{c|}{\textbf{Emulators}} & \multicolumn{2}{c}{\textbf{Tools}} \\
    \cmidrule(lr){1-1} \cmidrule(lr){2-7} \cmidrule(lr){8-9}
    \textbf{Name} & emulator\_x86 & Redroid\_arm & genymotion\_x86 & genymotion\_arm & emulator\_arm & mumu\_arm & VPBOX & \tool{} \\
    \textbf{Abnormal Rate (\%)} & 93 & 90 & 99 & 70 & 69 & 30 & 100 & 10 \\
    \bottomrule
  \end{tabular}
\end{table*}

\para{Evade apps' VM detection}
Table~\ref{tab:emulator_abnormal_rate} shows that \tool{} achieves the lowest abnormality rate among all evaluated environments. MuMu performs better than the other emulators, likely because it has been optimized against common environment checks for game workloads. However, these optimizations still fall short of the cross-layer consistency provided by \tool{}.

\para{Evade VM-detection heuristics}
We also test prior VM-detection heuristics drawn from earlier studies~\cite{Vidas2014SandboxDetection,bordoni2017mirage,petsas2014rage,sahin2018proteus,shi2019jekyll,zheng2018plugin,android_safetynet_attestation}. 
After removing obsolete checks that no longer apply to Android~14, we keep nine representative heuristic families. Table~\ref{tab:detection_coverage} shows that mainstream emulators and VPBOX remain detectable, while \tool{} performs best across the full suite.

\begin{table*}[!htbp]
  \centering
  \footnotesize
  \caption{Ad Appearance Rate of Different Emulators, Tools, and Real Device}
  \label{tab:emulator_ad_rate}
  \begin{tabular}{c|cccccc|cc|c}
    \toprule
    \textbf{Type} & \multicolumn{6}{c|}{\textbf{Emulators}} & \multicolumn{2}{c|}{\textbf{Tools}} & \textbf{Real Device} \\
    \cmidrule(lr){1-1} \cmidrule(lr){2-7} \cmidrule(lr){8-9} \cmidrule(lr){10-10}
    \textbf{Name} & emulator\_x86 & Redroid\_arm & genymotion\_x86 & genymotion\_arm & emulator\_arm & mumu\_arm & VPBOX & \tool{} & VIVO S17e \\
    \textbf{Ad Appearance Rate (\%)} & 0 & 4 & 0 & 0 & 12 & 26 & 0 & 100 & 100 \\
    \bottomrule
  \end{tabular}
\end{table*}

\para{Ad appearance rate}
Besides checking for app execution abnormalities, we consider the triggering of semi-drive-by ads an important metric, as it not only suggests successful evasion of VM detection by ad networks, but is also a prerequisite for subsequent fraudulent interactions.
To report this metric, we randomly sampled apps from Company-A's user complaint records that had been manually confirmed to display semi-drive-by ads.
We ran them in \tool{}, VPBOX, the mainstream emulators, as well as on the corresponding real device.
All environments used the same app version, network settings, and other configurations.
Each app was cold-launched 10 times, with each run lasting 1 minute and runs spaced 5 minutes apart. We counted any semi-drive-by ad shown during the observation window as a valid exposure.
Table~\ref{tab:emulator_ad_rate} shows that \tool{} consistently outperforms existing emulators and analysis frameworks. More importantly, the results suggest that bypassing some VM checks is not enough. Reliable exposure also requires a realistic usage context. For example, emulator\_arm reduces abnormal outcomes to some extent, but still exposes very few ads.

%

\subsubsection{\underline{Usage trace ablation detection}}
To study the effect of usage-trace synthesis, we run an ablation experiment in \tool{} with and without synthetic traces. We randomly sample 100 applications from all apps in Company-A's user complaint records that had previously been manually confirmed on real devices to exhibit \fraud{}.
Because triggering is stochastic, we repeatedly launch each app, starting from 5 runs and increasing the number of launches until the cumulative number of valid exposures stops growing. Table~\ref{tab:eval_summary} reports representative counts.

Without usage traces, the system saturates early at 18 valid exposures. With usage traces, it continues to uncover new cases and reaches 96 valid exposures by 75 launches. The gap shows that realistic traces are important for exposing behaviorally gated \fraud{}. Among the few remaining misses, three apps were blocked by updates and one failed because of ad-loading errors.

\subsubsection{\underline{Pipeline Analyzer Evaluation}}

\label{event_log_evaluation}

To assess whether our log analyzer can reliably identify high-confidence fraudulent logs, we select 100 benign logs and
100 advertising logs from preliminary experiments conducted before the large-scale measurement. These logs were independently labeled by two researchers; labeling details are provided in Appendix~\ref{piplinedetails}.

Table~\ref{tab:gpt_rule_expansion_metrics} shows that GPT-4o-assisted rule expansion improves high-confidence fraud detection from 79/100 to 91/100 fraudulent logs, while keeping the false positive rate on 100 benign logs at 0.0\%. This improves recall from 79.0\% to 91.0\% and F1 from 88.27\% to 95.29\%, with precision remaining at 100.0\%. Since all GPT-4o-proposed expressions are manually verified before being added to the deterministic rule dictionary, GPT-4o improves coverage without directly determining final labels. In total, the process surfaced 19 new advertisement-attribution fields.

We also provide prediction distribution and ablation analysis in Appendix~\ref{appendix:prediction} and \ref{appendix:ablation}.

\begin{table}[t]
\centering
\caption{Performance of High-Confidence Fraud Log Detection}
\label{tab:gpt_rule_expansion_metrics}
\scriptsize
\setlength{\tabcolsep}{3pt}
\resizebox{\columnwidth}{!}{%
\begin{tabular}{lccccccccc}
\toprule
Method & TP & FP & FN & TN & Prec. & Rec. & F1 & FPR & Acc. \\
\midrule
Expert rules only & 79 & 0 & 21 & 100 & 100.0\% & 79.0\% & 88.27\% & 0.0\% & 89.5\% \\
Expert rules + GPT-4o & 91 & 0 & 9 & 100 & 100.0\% & 91.0\% & 95.29\% & 0.0\% & 95.5\% \\
\bottomrule
\end{tabular}
\vspace{1pt}
}
\end{table}

\begin{table}[t]
\centering
\scriptsize
\caption{Ad Fraud Detection and Triggers in App Markets}
\label{tab:ad_fraud_summary}
\setlength{\tabcolsep}{3pt}
\renewcommand{\arraystretch}{1}
\begin{tabular}{cccccc}
\toprule
\textbf{Market} & \textbf{Tested} & \textbf{Fraud} & \textbf{Rate} & \textbf{Sensor} & \textbf{Tap/Swipe/Scroll} \\
\midrule
VIVO   & 5,262  & 261 & 4.96 & 249 & 12 \\
XIAOMI & 5,998  & 128 & 2.13 & 47  & 81 \\
WDJ    & 11,682 & 198 & 1.70 & 180 & 18 \\
YYB    & 8,881  & 191 & 2.15 & 168 & 13 \\
\bottomrule
\end{tabular}
\end{table}

\section{Large-Scale Measurement}
\label{sec-measurement}

\subsection{Data Collection}
\label{subsec:data_collection}

Independent industry reports indicate that advertising fraud rates and absolute fraud volume are markedly higher in China than in other regions~\cite{ppcshield2025,cheq2021,groupm2019}. For example, GroupM estimates that China accounts for a dominant share of global advertising fraud~\cite{groupm2019}. Motivated by this context, we scope our study to the Chinese Android ecosystem.

We crawled all \emph{non-game} and \emph{popular} categories from four major Chinese Android markets:
XIAOMI, VIVO, YYB and WDJ, obtaining 5{,}480, 6{,}026, 11{,}928, and 9{,}324 applications, respectively, for a total of 32{,}758 APKs.
Because ad fraud requires network access, we used Androguard~\cite{Androguard_Docs_2025} to filter out packages without Internet permission, and we excluded apps that proved incompatible at runtime when executed in \tool{}. 
In total, we discarded 104 apps for lacking Internet permissions and 831 for incompatibility, leading to the final test set of \textbf{31{,}823} apps.

All experiments were executed on a dedicated cluster of \textbf{15} Alibaba Cloud ECS \texttt{ecs.g8y.16xlarge} instances (ARM, 64 vCPUs, 256\,GiB RAM each) and one Apple Mac M1 Pro host (32\,GiB RAM, 1\,TB storage). We orchestrated approximately \textbf{200} instrumented Android virtual devices across this cluster.

We begin with the overall prevalence picture, then break down trigger types, target categories, regional differences, and cross-market behavioral variation. This ordering helps separate what is broadly common from what changes with market structure or device context.

\subsection{How prevalent are \fraud{} across markets?}

We executed each app using \tool{} for 24 hours under a uniform schedule: launch the app, return to the homepage, and repeat every 5 minutes. \tool{} captures and classifies DeepLink requests and launch events attributable to \fraud{}. Table~\ref{tab:ad_fraud_summary} summarizes prevalence and trigger breakdown by market. We found 261, 128, 198, and 191 apps with \fraud{} in the VIVO, XIAOMI, WDJ, and YYB app markets, respectively. The ratios of apps with splash ad fraud in each app store are 4.96\%, 2.13\%, 1.70\%, and 2.15\%.

We observe non-trivial prevalence in all four markets, with VIVO exhibiting the highest rate (4.96\%). 
Although category taxonomies vary by market, we manually harmonized categories to enable comparison. In our preliminary analysis, we found that direct-launch behaviors only occur in tap/scroll/swipe-type ads; we therefore group them into a single interaction category. Table~\ref{tab:category-four-markets} reports category-level statistics per market.
Across markets, \textit{tools}, \textit{photography}, and \textit{lifestyle}—three categories closely tied to daily use—are consistently overrepresented among \fraudname{} ad targets.

\begin{figure}[b]
    \centering
    \vspace{5pt} 
\includegraphics[width=\linewidth,height=\textheight,keepaspectratio,alt={}]{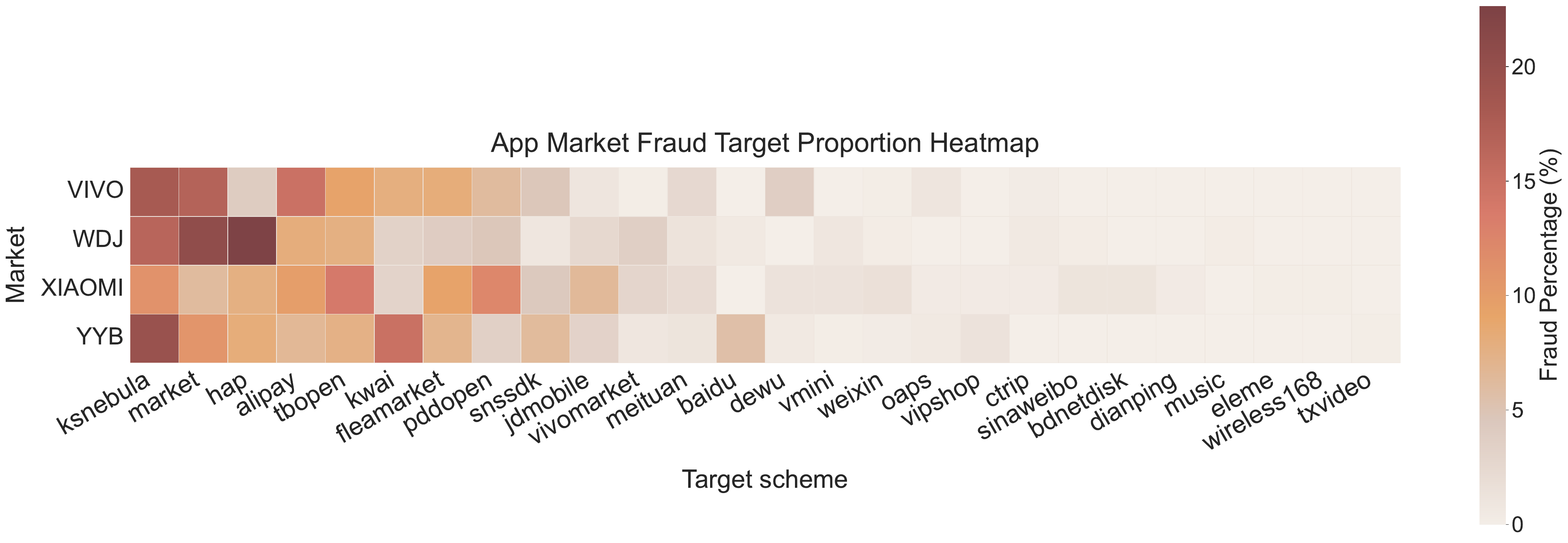}
    \caption{Affected App Distribution in 4 App Markets} 
    \label{fig:heatmapoftarget}
\end{figure}

\subsection{Sensor-triggered vs.\ tap/scroll/swipe}

To disentangle sensor-triggered \fraud{} from visually induced variants (e.g., deceptive tap/scroll/swipe), we re-ran all flagged apps with motion sensors disabled and otherwise identical conditions. We treat the two modes as mutually exclusive at runtime for a given app configuration. Table~\ref{tab:ad_fraud_summary} reports the breakdown.

\definecolor{colT}{RGB}{183,220,240}    
\definecolor{colTP}{RGB}{248,207,193}   
\definecolor{colS}{RGB}{199,233,197}    
\definecolor{colSP}{RGB}{227,214,235}   
\definecolor{colD}{RGB}{247,230,191}    
\definecolor{colDP}{RGB}{202,234,234}   

\newcommand{\CellT}[2]{\cellcolor{colT!#1}{#2}}    
\newcommand{\CellTP}[2]{\cellcolor{colTP!#1}{#2}}  
\newcommand{\CellS}[2]{\cellcolor{colS!#1}{#2}}    
\newcommand{\CellSP}[2]{\cellcolor{colSP!#1}{#2}}  
\newcommand{\CellD}[2]{\cellcolor{colD!#1}{#2}}    
\newcommand{\CellDP}[2]{\cellcolor{colDP!#1}{#2}}  

\begin{table*}[htbp]
\centering
\caption{Category-wise Statistics for Four App Markets}
\label{tab:category-four-markets}

\begin{subtable}[t]{0.48\textwidth}
\centering
\caption{Category-wise Statistics for VIVO}
\begin{tabular}{lrrrrrr}
\toprule
\multirow{2}{*}{\textbf{Category}} & \multicolumn{6}{c}{\textbf{VIVO}} \\
\cmidrule(lr){2-7}
 & \textbf{T} & \textbf{TP(\%)} & \textbf{S} & \textbf{SP(\%)} & \textbf{D} & \textbf{DP(\%)} \\
\midrule
car           & \CellT{12}{11} & \CellTP{12}{4.21}  & \CellS{11}{10} & \CellSP{11}{4.02}  & \CellD{25}{1} & \CellDP{25}{8.33} \\
edu           & \CellT{4}{4}  & \CellTP{4}{1.53}  & \CellS{4}{4}  & \CellSP{4}{1.61}  & \CellD{0}{0} & \CellDP{0}{0} \\
finance       & \CellT{8}{7}  & \CellTP{8}{2.68}  & \CellS{6}{5}  & \CellSP{6}{2.01}  & \CellD{50}{2} & \CellDP{50}{16.67} \\
health        & \CellT{28}{26} & \CellTP{27}{9.96} & \CellS{26}{24} & \CellSP{26}{9.64} & \CellD{50}{2} & \CellDP{50}{16.67} \\
lifestyle     & \CellT{16}{15} & \CellTP{16}{5.75} & \CellS{16}{15} & \CellSP{16}{6.02} & \CellD{0}{0} & \CellDP{0}{0} \\
office        & \CellT{2}{2}  & \CellTP{2}{0.77}  & \CellS{2}{2}  & \CellSP{2}{0.80}  & \CellD{0}{0} & \CellDP{0}{0} \\
photography   & \CellT{34}{33} & \CellTP{35}{12.64} & \CellS{34}{33} & \CellSP{36}{13.25} & \CellD{0}{0} & \CellDP{0}{0} \\
reading       & \CellT{8}{7}  & \CellTP{8}{2.68}  & \CellS{7}{6}  & \CellSP{7}{2.41}  & \CellD{25}{1} & \CellDP{25}{8.33} \\
shopping      & \CellT{2}{2}  & \CellTP{2}{0.77}  & \CellS{2}{2}  & \CellSP{2}{0.80}  & \CellD{0}{0} & \CellDP{0}{0} \\
social        & \CellT{34}{32} & \CellTP{33}{12.26} & \CellS{34}{31} & \CellSP{34}{12.45} & \CellD{25}{1} & \CellDP{25}{8.33} \\
tools         & \CellT{100}{95} & \CellTP{100}{36.40} & \CellS{100}{91} & \CellSP{100}{36.55} & \CellD{100}{4} & \CellDP{100}{33.30} \\
entertainment & \CellT{28}{27} & \CellTP{28}{10.35} & \CellS{28}{26} & \CellSP{28}{10.44} & \CellD{25}{1} & \CellDP{25}{8.33} \\
\bottomrule
\end{tabular}
\end{subtable}
\hfill
\begin{subtable}[t]{0.48\textwidth}
\centering
\caption{Category-wise Statistics for XIAOMI}
\begin{tabular}{lrrrrrr}
\toprule
\multirow{2}{*}{\textbf{Category}} & \multicolumn{6}{c}{\textbf{XIAOMI}} \\
\cmidrule(lr){2-7}
 & \textbf{T} & \textbf{TP(\%)} & \textbf{S} & \textbf{SP(\%)} & \textbf{D} & \textbf{DP(\%)} \\
\midrule
edu           & \CellT{81}{21} & \CellTP{81}{16.41} & \CellS{100}{15} & \CellSP{100}{31.91} & \CellD{24}{6} & \CellDP{24}{7.41} \\
finance       & \CellT{4}{1} & \CellTP{4}{0.78} & \CellS{0}{0} & \CellSP{0}{0} & \CellD{4}{1} & \CellDP{4}{1.23} \\
health        & \CellT{15}{4} & \CellTP{15}{3.12} & \CellS{0}{0} & \CellSP{0}{0} & \CellD{16}{4} & \CellDP{16}{4.94} \\
lifestyle     & \CellT{88}{23} & \CellTP{88}{17.96} & \CellS{60}{9} & \CellSP{60}{19.15} & \CellD{56}{14} & \CellDP{56}{17.28} \\
office        & \CellT{15}{4} & \CellTP{15}{3.12} & \CellS{20}{3} & \CellSP{20}{6.38} & \CellD{4}{1} & \CellDP{4}{1.23} \\
photography   & \CellT{35}{9} & \CellTP{35}{7.03} & \CellS{27}{4} & \CellSP{27}{8.51} & \CellD{20}{5} & \CellDP{20}{6.17} \\
reading       & \CellT{27}{7} & \CellTP{27}{5.47} & \CellS{33}{5} & \CellSP{33}{10.64} & \CellD{8}{2} & \CellDP{8}{2.47} \\
shopping      & \CellT{15}{4} & \CellTP{15}{3.12} & \CellS{20}{3} & \CellSP{20}{6.38} & \CellD{4}{1} & \CellDP{4}{1.23} \\
social        & \CellT{19}{5} & \CellTP{19}{3.91} & \CellS{0}{0} & \CellSP{0}{0} & \CellD{20}{5} & \CellDP{20}{6.17} \\
tools         & \CellT{100}{26} & \CellTP{100}{20.32} & \CellS{7}{1} & \CellSP{7}{2.13} & \CellD{100}{25} & \CellDP{100}{30.87} \\
entertainment & \CellT{92}{24} & \CellTP{92}{18.74} & \CellS{47}{7} & \CellSP{47}{14.89} & \CellD{68}{17} & \CellDP{68}{20.98} \\
\bottomrule
\end{tabular}
\end{subtable}

\vspace{1em}

\begin{subtable}[t]{0.48\textwidth}
\centering
\caption{Category-wise Statistics for WDJ}
\begin{tabular}{lrrrrrr}
\toprule
\multirow{2}{*}{\textbf{Category}} & \multicolumn{6}{c}{\textbf{WDJ}} \\
\cmidrule(lr){2-7}
 & \textbf{T} & \textbf{TP(\%)} & \textbf{S} & \textbf{SP(\%)} & \textbf{D} & \textbf{DP(\%)} \\
\midrule
finance       & \CellT{7}{3} & \CellTP{7}{1.52} & \CellS{7}{3} & \CellSP{7}{1.67} & \CellD{0}{0} & \CellDP{0}{0} \\
office        & \CellT{34}{14} & \CellTP{34}{7.07} & \CellS{34}{14} & \CellSP{34}{7.78} & \CellD{0}{0} & \CellDP{0}{0} \\
photography   & \CellT{100}{41} & \CellTP{100}{20.71} & \CellS{100}{41} & \CellSP{100}{22.78} & \CellD{0}{0} & \CellDP{0}{0} \\
reading       & \CellT{51}{21} & \CellTP{51}{10.61} & \CellS{44}{18} & \CellSP{44}{10} & \CellD{75}{3} & \CellDP{75}{16.67} \\
shopping      & \CellT{5}{2} & \CellTP{5}{1.01} & \CellS{5}{2} & \CellSP{5}{1.11} & \CellD{0}{0} & \CellDP{0}{0} \\
social        & \CellT{12}{5} & \CellTP{12}{2.53} & \CellS{5}{2} & \CellSP{5}{1.11} & \CellD{75}{3} & \CellDP{75}{16.67} \\
sports        & \CellT{15}{6} & \CellTP{15}{3.03} & \CellS{7}{3} & \CellSP{7}{1.67} & \CellD{75}{3} & \CellDP{75}{16.67} \\
entertainment & \CellT{37}{15} & \CellTP{37}{7.58} & \CellS{27}{11} & \CellSP{27}{6.11} & \CellD{100}{4} & \CellDP{100}{22.22} \\
tools         & \CellT{66}{27} & \CellTP{66}{13.64} & \CellS{59}{24} & \CellSP{59}{13.33} & \CellD{75}{3} & \CellDP{75}{16.67} \\
edu           & \CellT{59}{24} & \CellTP{59}{12.12} & \CellS{56}{23} & \CellSP{56}{12.78} & \CellD{25}{1} & \CellDP{25}{5.56} \\
lifestyle     & \CellT{98}{40} & \CellTP{98}{20.21} & \CellS{95}{39} & \CellSP{95}{21.67} & \CellD{25}{1} & \CellDP{25}{5.56} \\
\bottomrule
\end{tabular}
\end{subtable}
\hfill
\begin{subtable}[t]{0.48\textwidth}
\centering
\caption{Category-wise Statistics for YYB}
\begin{tabular}{lrrrrrr}
\toprule
\multirow{2}{*}{\textbf{Category}} & \multicolumn{6}{c}{\textbf{YYB}} \\
\cmidrule(lr){2-7}
 & \textbf{T} & \textbf{TP(\%)} & \textbf{S} & \textbf{SP(\%)} & \textbf{D} & \textbf{DP(\%)} \\
\midrule
car           & \CellT{7}{5} & \CellTP{7}{2.62} & \CellS{6}{4} & \CellSP{6}{2.38} & \CellD{20}{1} & \CellDP{19}{4.35} \\
finance       & \CellT{3}{2} & \CellTP{3}{1.05} & \CellS{1}{1} & \CellSP{1}{0.6} & \CellD{20}{1} & \CellDP{19}{4.35} \\
health        & \CellT{5}{4} & \CellTP{5}{2.09} & \CellS{3}{2} & \CellSP{3}{1.19} & \CellD{40}{2} & \CellDP{39}{8.7} \\
office        & \CellT{12}{9} & \CellTP{12}{4.71} & \CellS{13}{9} & \CellSP{13}{5.36} & \CellD{0}{0} & \CellDP{0}{0} \\
photography   & \CellT{32}{24} & \CellTP{32}{12.57} & \CellS{28}{20} & \CellSP{28}{11.9} & \CellD{80}{4} & \CellDP{80}{17.39} \\
reading       & \CellT{9}{7} & \CellTP{9}{3.66} & \CellS{7}{5} & \CellSP{7}{2.98} & \CellD{40}{2} & \CellDP{39}{8.7} \\
shopping      & \CellT{8}{6} & \CellTP{8}{3.14} & \CellS{7}{5} & \CellSP{7}{2.98} & \CellD{20}{1} & \CellDP{19}{4.35} \\
social        & \CellT{4}{3} & \CellTP{4}{1.57} & \CellS{4}{3} & \CellSP{4}{1.79} & \CellD{0}{0} & \CellDP{0}{0} \\
tools         & \CellT{42}{32} & \CellTP{42}{16.75} & \CellS{38}{27} & \CellSP{38}{16.08} & \CellD{100}{5} & \CellDP{100}{21.74} \\
edu           & \CellT{7}{5} & \CellTP{7}{2.61} & \CellS{7}{5} & \CellSP{7}{2.98} & \CellD{0}{0} & \CellDP{0}{0} \\
lifestyle     & \CellT{24}{18} & \CellTP{24}{9.42} & \CellS{21}{15} & \CellSP{21}{8.94} & \CellD{60}{3} & \CellDP{60}{13.05} \\
entertainment & \CellT{100}{76} & \CellTP{100}{39.79} & \CellS{100}{72} & \CellSP{100}{42.85} & \CellD{80}{4} & \CellDP{80}{17.39} \\
\bottomrule
\end{tabular}
\end{subtable}

\vspace{0.5ex}
\begin{minipage}{\linewidth}
\footnotesize
\textit{Note.} T: total Semi-Drive-By apps observed; TP: share of Semi-Drive-By apps attributable to the category; S: sensor-triggered Semi-Drive-By apps; SP: share of sensor-triggered Semi-Drive-By apps attributable to the category; D: tap/scroll/swipe Semi-Drive-By apps; DP: share of tap/scroll/swipe Semi-Drive-By apps attributable to the category.
\end{minipage}
\end{table*}

Sensor-triggered \fraud{} dominate in three markets (VIVO, WDJ, YYB). XIAOMI shows the opposite balance. We hypothesize this difference is influenced by store-level listing and policy controls; a causal analysis is outside the scope of our measurement.

\subsection{Which targets are affected?}

We categorize impacted targets from our large-scale runs and visualize them in Figure~\ref{fig:heatmapoftarget}. The most frequently implicated families are leading e-commerce and short-video platforms (e.g., Alibaba, Ant Group, PDD, JD, Kuaishou), which typically allocate large user-acquisition budgets.

\para{Popular apps}
From the heatmap, we observe that the target applications of \fraud{} are mainly concentrated on leading e-commerce and short video platforms, such as PDD Holdings Inc., Alibaba, Ant Finance, Kuaishou Technology, and Jingdong. The apps from these companies are very popular among Chinese users. We observe that splash ads are more prevalent in e-commerce and short-video applications. This concentration of targets may be related to the intensive advertising activities commonly observed on these platforms. Recent industry trends also suggest increasing investment in digital advertising in these domains. However, we do not claim a causal relationship here. From the perspective of ad placement providers, e-commerce applications such as Taobao, Pinduoduo, and Xianyu, as well as short video applications such as Kuaishou and its international version, typically invest heavily in user acquisition and traffic conversion. This makes them primary targets for fraudsters, who can simulate real traffic through fake launch operations and thereby illegally obtain advertising revenue shares or inflate performance indicators.

\para{Super apps}
From the logs we collected, we found that \fraud{} also target super apps such as QuickApp~\cite{8818418} and WeChat Mini Programs~\cite{hao2018analysis}. Specifically, some fraud instances do not launch standalone apps but instead open advertisement pages within mini programs supported by QuickApp or WeChat. As shown in Table~\ref{tab:app_markets}, across the four app markets, the most prominent target types primarily include QuickApps and app-store download targets. Therefore, we further counted and categorized the mini programs launched via QuickApps and the app-store download targets, and we arranged the regional comparison in the same cross-column row for a compact view. We observe that QuickApps are mainly concentrated in the reading and shopping categories. For further analysis, we manually examined the DeepLink URIs of QuickApps and found that, for reading-related QuickApps, \textit{fraud} not only launches the reading mini program itself but also directs traffic to specific novels within it for additional promotion.

\begin{figure*}[t]
    \centering
    \begin{subfigure}[t]{0.33\textwidth}
        \centering
        \includegraphics[width=\linewidth,keepaspectratio]{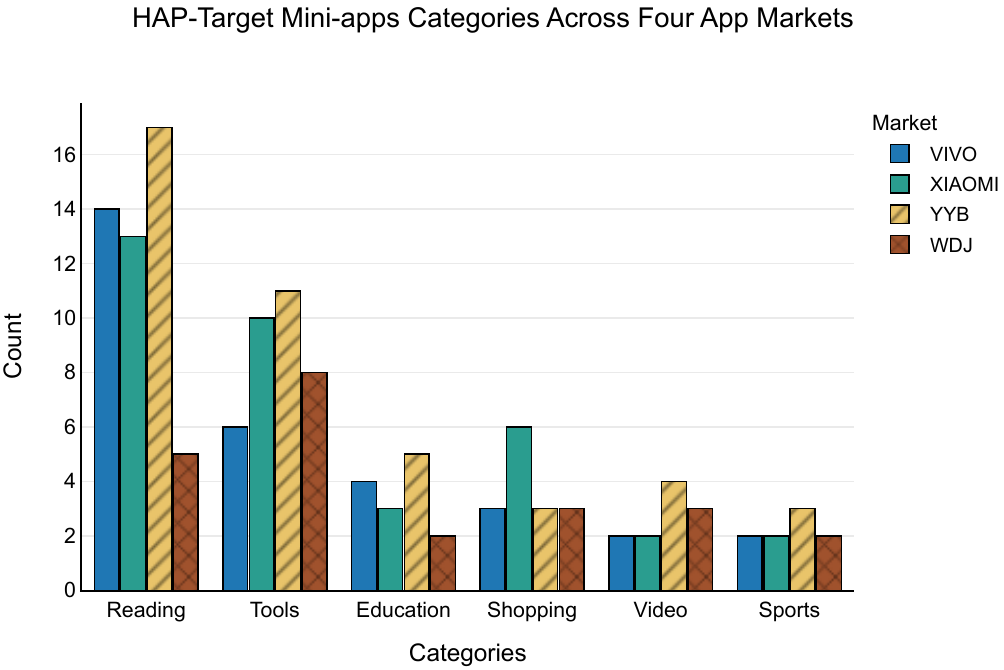}
        \caption{HAP categories}
    \end{subfigure}
    \hfill
    \begin{subfigure}[t]{0.33\textwidth}
        \centering
        \includegraphics[width=\linewidth,keepaspectratio]{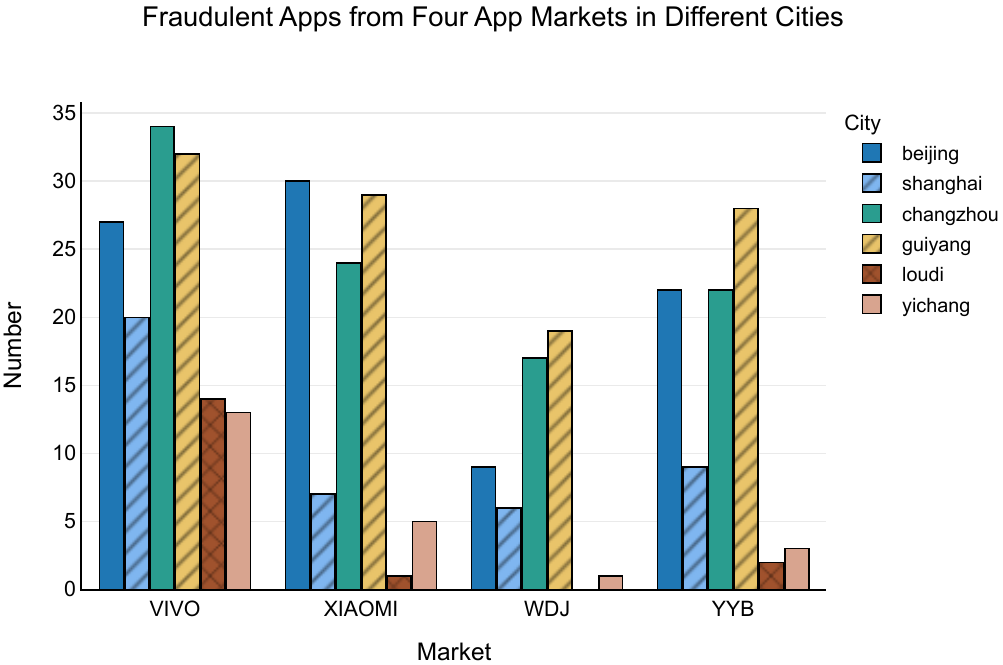}
        \caption{Regional comparison}
    \end{subfigure}
    \hfill
    \begin{subfigure}[t]{0.32\textwidth}
        \centering
        \includegraphics[width=\linewidth,keepaspectratio]{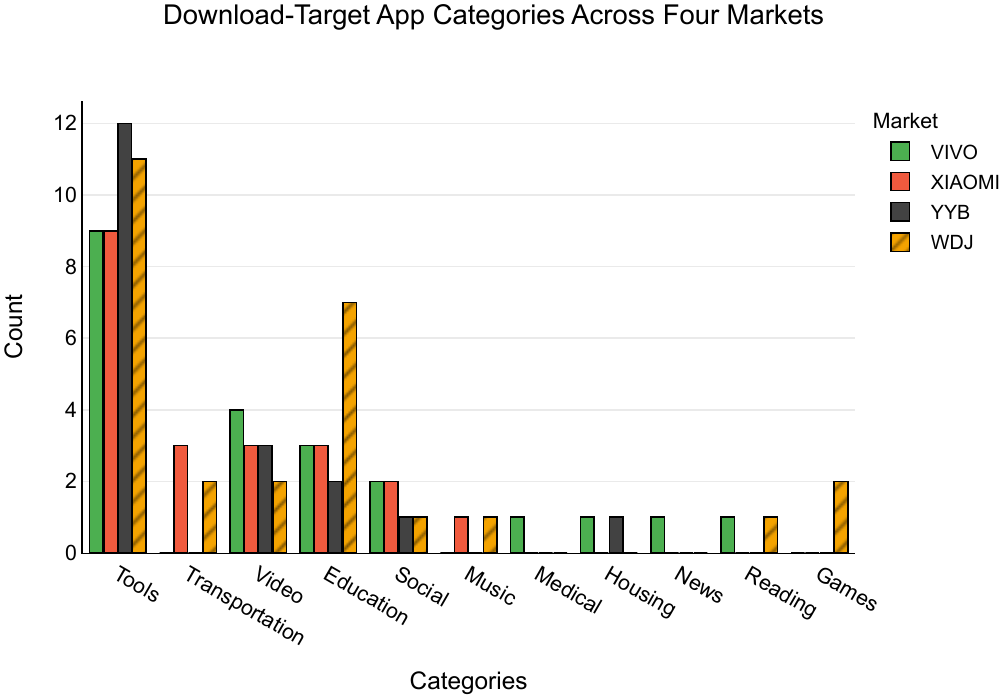}
        \caption{Market categories}
    \end{subfigure}
    \caption{Comparison of special target categories and regional distribution across four app markets.}
    \label{fig:hap_market_category}
\end{figure*}

\para{App-market app page}
From Table~\ref{tab:app_markets}, we observe that among these download-type targets, the promoted downloads are mainly concentrated in life-related utility and education applications. In addition, we found that some applications can be forcibly downloaded and installed through the app store.
By using the app store as an intermediary to initiate a DeepLink request, the store can verify whether the target app is installed. If the app is installed, it will be launched. From the perspective of the launched app, the DeepLink request originates from the app market, which is considered a trusted platform. As a result, such requests may undergo less verification.

\subsection{Economic-region behavioral differences}
\begin{table}[b]
\centering
\caption{Market-wise Target Statistics}
\label{tab:app_markets}
\resizebox{0.8\columnwidth}{!}{%
\begin{tabular}{c|cccccc}
\toprule
\multirow{2}{*}{Market} & \multicolumn{6}{c}{Target} \\
\cmidrule(lr){2-7}
 & Hap & Market & Oaps & VivoMarket & VMini & Weixin \\
\midrule
XIAOMI & 37 & 21 & 0 & 7 & 5 & 1 \\
WDJ    & 23 & 27 & 0 & 5 & 1 & 0 \\
YYB    & 43 & 19 & 3 & 6 & 1 & 1 \\
VIVO   & 31 & 22 & 1 & 3 & 2 & 1 \\
\bottomrule
\end{tabular}
}
\begin{minipage}{\columnwidth}
\vspace{5pt}
\footnotesize
\textit{Note.} Hap denotes QuickApp-based miniapps. Market denotes app-store download pages. Oaps denotes the app-store pages. VivoMarket denotes Vivo app-store pages. VMini denotes Vivo quick apps. Weixin denotes WeChat miniapps.
\end{minipage}
\end{table}

We further conducted regional comparison experiments. We randomly sampled 1,000 apps from the four app markets and performed comparative analyses across regions. Specifically, based on the GDP classification of Chinese cities, we selected first-tier (Beijing, Shanghai), second-tier (Changzhou, Guiyang), and third-tier (Yichang, Loudi) cities—two from each category, for a total of six cities—corresponding to six groups of virtual machines (each group comprising 100 devices). We then conducted 24-hour continuous app-launch detection experiments.
The regional panel in Figure~\ref{fig:hap_market_category}(b) indicates that \fraud{} are detected most frequently in second-tier cities, followed by first-tier cities, and are much less common in third-tier cities, where the detection volume is significantly lower. This suggests that ad networks may concentrate their placement efforts primarily in second-tier cities.

Regionally, \fraud{} in Beijing are dominated by Kuaishou-related and download-type apps, whereas in second-tier cities they are mainly associated with the Alibaba app ecosystem and cover the broadest range of target apps. Third-tier cities exhibit the least diversity. In Shanghai, as a representative first-tier city, fraud activity types are more evenly distributed across the four app markets.

\subsection{Different behavior from the same app in different app markets}

We found that the same app obtained from different app markets can exhibit different \fraudname{} behaviors.
Specifically, \fraud{} may appear in one app market but not in another. In total, 57 apps displayed cross-market discrepancies: 42 differed between two markets and 15 differed across three markets, while none differed across all four markets. Among these cases, 18 apps in the two-market group and 8 apps in the three-market group shared the same version number, indicating that version drift alone does not explain the behavioral differences.

Although in our dataset the same application may appear in different versions across various app markets—suggesting that version differences could explain variations in \fraud{} behavior—the counts show that this is not the case. Even identical versions can exhibit different fraud behaviors.

To identify the underlying causes, we modified LibChecker\cite{LibChecker} to automatically analyze apps with the same version obtained from different markets. The tool generates a JSON file listing library versions and file sizes, which we compare by library name and size to identify differences. We found two cases in which the apps contained different SDKs or used different ad push services. This may be one of the reasons for the variation in fraud behavior.

\subsection{Ad Networks or App Developers?}

\para{Attribution methodology}
To identify the true source of the fraud behavior, we randomly selected 100 apps detected by \tool{} and conducted manual reverse engineering with three independent security experts. Each expert traced the origin of the suspicious logic following a consistent procedure: (1) locate code segments acquiring and processing sensor data, (2) identify motion-triggered decision logic and ad-related callbacks, and (3) iteratively trace call graphs or conduct large-scale pattern searches to attribute the behavior to a specific SDK. Across all 100 cases, the fraud code was located within advertising SDKs, not within developer-authored components. 
%
%
We further reviewed the corresponding SDK documentation and found no evidence that app developers could control any parameters related to the ad-fraud behaviors.
This indicates that the ad networks themselves, rather than app developers, are responsible for implementing and distributing the fraudulent logic.

\para{Economic and structural motivations}
Prior work~\cite{crussell2014madfraud,shekhar2012adsplit,cma2020intermediation} has characterized ad networks as intermediaries between advertisers and publishers, with control over both ad delivery and ad measurement and reporting~\cite{cma2020intermediation,cma2020measurement}.
This privileged position creates an opportunity and, where advertiser payments depend on reported engagement, a potential economic incentive to inflate engagement metrics.
%
This problem is reflected not only in prior regulatory and industry studies documenting limited transparency and traceability in advertising~\cite{cma2020measurement,cma2020intermediation,isba2020transparency,ana2023transparency}, but also in our study, which offers concrete evidence that ad networks manipulate interaction metrics to increase revenue.
Independent app developers, by contrast, have less incentive to engage in this type of fraud. They would bear legal and reputational risks while receiving only a portion of the additional advertising revenue. Ad networks, however, benefit directly from higher reported engagement and are better positioned, both legally and operationally, to manage potential disputes over advertising measurements.



\para{Technical rationale for \fraud{}}
Unlike clicks or swipes, which can be recorded as discrete interaction events, motion-triggered splash ads rely on continuous sensor data that are often not routinely retained due to performance and privacy concerns (as indicated by the privacy policies of several ad SDKs~\cite{OctopusSDKPrivacy,HonorAdsSDKPrivacy,ToBidPrivacy}, which state that sensor data are used only locally on the device to trigger ads and are not uploaded to their servers).
%
%
This creates an auditing blind spot for advertisers, who may lack the sensor evidence needed to determine whether a reported motion-triggered activation reflects genuine user intent. As a result, unintended or manipulated activations may still appear as legitimate engagement in ad reporting.
This limitation also makes semi-drive-by ads difficult to identify from both the user and advertiser sides. Users may attribute accidental launches to normal device handling and therefore rarely report them, while advertisers lack sufficient evidence to challenge the reported metrics. 
Together, these properties make semi-drive-by ads a low-risk way to inflate click-through and conversion rates without overtly falsifying reports.

\section{Discussion}
\label{discussion}

Our findings have two implications. First, semi-drive-by ad fraud requires treating ad SDKs, rather than only app developers, as part of the adversarial surface. Second, our measurement should be interpreted within the scope and limits of the study design.

\subsection{Implications}

Our findings reveal that ad network SDKs, rather than app developers, can operate as the primary adversarial agents in modern mobile advertising ecosystems. This shift in threat boundary implies that existing fraud detection frameworks, which primarily monitor developer behavior or app-side anomalies, are no longer sufficient. Industry and regulators should broaden their oversight to include ad network auditing and SDK-level accountability. Our automated, environment-resilient measurement tool, \tool{}, can play a central role in enabling independent verification of ad delivery integrity and supporting evidence-based compliance actions.

From a broader perspective, the prevalence of semi-drive-by ads underscores the fragility of trust in data-driven monetization systems. Advertisers bear direct financial losses, while users experience covert manipulation of their interactions. Addressing these systemic risks requires stronger transparency mandates on SDK behavior, standardized reporting of sensor-triggered events, and collaborative auditing between advertisers, regulators, and independent researchers. Our work provides a technical foundation for these accountability mechanisms.

\subsection{Limitation}

Our measurement focuses on Android splash ads in four Chinese app markets and uses hardened emulation with synthesized motion traces. These choices bound external validity (e.g., other regions, iOS, and non-splash formats) and may leave residual gaps in trigger realism. The 24-hour, 5-minute launch schedule standardizes coverage but can miss time-gated, quota-based, or A/B–rolled activations; manual category harmonization may introduce label noise. Detection assumes mutual exclusivity between sensor-triggered and tap/scroll/swipe variants under a given configuration, and relies on DeepLink/event-log signatures; adaptive SDK behavior or obfuscated telemetry could yield false negatives or, less likely, flag aggressive but legitimate motion UX.
%
%
%
We report market and regional differences without claiming causality; unobserved confounders (campaign mix, moderation actions) may contribute. Ethical constraints (no production user data, redacted logs) and limited access to proprietary billing ledgers restrict end-to-end spend reconciliation beyond the \textit{Company~A} case. Finally, disclosure may prompt adversarial adaptation (e.g., tighter VM checks, randomized thresholds), so our prevalence estimates reflect a specific study window rather than a static equilibrium.

To detect fraud in shake-to-trigger ads, we adopt threshold values defined by the industry advertising standard TC260~\cite{TC2602025ShakeAds}. 
These thresholds are intended to prevent ads from being triggered by ordinary user behaviors, such as walking, riding in a vehicle, or picking up or putting down a device.
While these thresholds help identify a number of semi-drive-by ads, their selection introduces a trade-off that our study does not fully explore: lower thresholds may lead to more false positives, whereas higher thresholds may lead to more false negatives. 
In addition, ad networks may use adaptive strategies that dynamically adjust their triggering conditions, which could reduce the effectiveness of static thresholds. 
Therefore, future research should investigate these threshold-selection trade-offs and further examine whether advanced ad fraud requires adaptive detection methods and, if so, what types of adaptive methods are needed.
Furthermore, the validation of semi-drive-by ads is based on these threshold values and the authors' empirical judgments. %
There is a general lack of validation from end users' perspectives, i.e., whether users actually perceive such ads as problematic, which calls for future user studies to validate the practical impact and perceived intrusiveness of these threats.

\section{Related Work}
\label{related_work}

\para{Ad fraud detection}
%
There has been an extensive body of prior research on detecting ad fraud, which generally falls into three categories.
First, bot- and click-farm-oriented studies detect fraudulent traffic by correlating clicks with genuine input events, testing browser capabilities and post-click behavior, or identifying repeated click-stream patterns~\cite{adsherlock,fcfraud,miller2011whats,2012click,xu2014click,zhang2008detecting}.
Second, placement-oriented systems such as DECAF~\cite{liu2014decaf} and FraudDroid~\cite{DBLP:conf/sigsoft/DongWLGBLXK18} detect deceptive ad layouts, overlays, background pop-ups, and UI-triggered abuses.
Third, programmatic-click studies identify automatic redirections, background ad requests, synthetic clicks, and replayed click patterns~\cite{madlife,crussell2014madfraud,clicktok}.
This study differs from prior work in two aspects.
First, a large portion of prior work considers app developers (or publishers) as the adversary~\cite{adsherlock,fcfraud,miller2011whats,xu2014click,2012click,zhang2008detecting,liu2014decaf,DBLP:conf/sigsoft/DongWLGBLXK18,crussell2014madfraud,madlife,clicktok}, while \tool{} focuses on investigating how ad networks become the major threat actors in splash ads.
Second, while a recent study attributed ad fraud to in-app modules (including third-party ad libraries)~\cite{kim2021abuser}, it did not systematically characterize how ad networks convert incidental or indirect interactions (e.g., motion- or gesture-driven triggers) into billable engagement, nor did it provide a distinct understanding of the splash ad ecosystem.

\para{Evasion behaviors}
In the ongoing arms race between attacks and defenses, adversaries often use environment-aware evasion techniques, in which they first identify the software execution environment and then suppress or alter malicious behavior when virtualization, emulation, sandboxing, or analysis instrumentation is detected.
For example, Vidas et al.~\cite{Vidas2014SandboxDetection} studied how malware evades dynamic analysis by detecting emulators.
Similar evasive behaviors were also observed in the ads ecosystem where ad fraud is hidden when software (or apps) run in emulators or sandboxes~\cite{DBLP:conf/sigsoft/DongWLGBLXK18,DBLP:conf/ccs/ZhuMHZXZ21}. 
To better analyze evasive behaviors, two major technical lines have been investigated.
First, prior work has explored how to detect evasive behaviors, e.g., through static trigger analysis~\cite{fratantonio2016triggerscope}, learning-based detection of hidden sensitive operations~\cite{pan2017dark,samhi2022difuzer}, and dynamic or hybrid analysis of environment-dependent behaviors~\cite{kirat2014barecloud,afonso2018lumus}.
Second, researchers have explored hardened virtualization and sandboxing techniques such that the execution environment exposes fewer detectable traces and more closely approximates real-device behavior~\cite{andrus2011cells,song2021vpbox,9401988}.
In general, \tool{} is an instance of the second technical line.
However, by combining environment-resilient honeypot designs, such as porting real-device profiles and synthesizing usage traces, \tool{} substantially enhances the appearance of semi-drive-by ads (as confirmed by the comparative evaluation in Section~\ref{subsec-validation}), thereby enabling reliable observation and attribution of semi-drive-by behaviors at scale.



\section{Conclusion}
\label{conc}

We present the first systematic study of semi-drive-by splash ad fraud originating from ad network SDKs. We empirically show that this behavior is widespread, evasive, and economically motivated. 

To expose it, we design \tool{}, an automated honeypot framework that induces and identifies such fraudulent activations at scale. Our results reveal a new platform-level threat in mobile advertising and provide a practical basis for future detection and accountability.

\section*{Acknowledgements}
We thank Dr. Qin Wang (senior research scientist at CSIRO, Australia) for his valuable assistance in improving the writing of this paper. His editing and polishing greatly improved the overall presentation of the manuscript. We also thank Lin Jiang, the author of VPBOX, for his help with the comparative experiments. Finally, we thank Yuqian Jiang, a security engineer at PDD, for his communication and support during the vulnerability reporting and disclosure process.



\bibliographystyle{unsrt}
\small{\bibliography{ref}}

\appendix

\input{section/ethics}


\section{Supplementary Implementation Details}
\label{appendix:impl_supp}

\begin{table}[!htbp]
\centering
\caption{APIs, categories, and corresponding key system calls.}
\label{tab:file_calls}
\scriptsize
\setlength{\tabcolsep}{5pt}
\renewcommand{\arraystretch}{1.0}
\begin{tabularx}{\linewidth}{@{}p{0.24\linewidth} p{0.16\linewidth} X@{}}
\toprule
API & Category & Key System Call \\
\midrule
\texttt{faccessat}  & Native  & \texttt{faccessat} \\
\texttt{stat64}     & Native  & \texttt{newfstatat} \\
\texttt{lstat}      & Native  & \texttt{newfstatat} \\
\texttt{lstat64}    & Native  & \texttt{newfstatat} \\
\texttt{statfs}     & Native  & \texttt{statfs} \\
\texttt{fstatfs}    & Native  & \texttt{statfs} \\
\texttt{fstat}      & Native  & \texttt{openat} \\
\texttt{fstat64}    & Native  & \texttt{openat} \\
\texttt{fstatat}    & Native  & \texttt{newfstatat} \\
\texttt{open}       & Native  & \texttt{openat} \\
\texttt{openat}     & Native  & \texttt{openat} \\
\texttt{fopen}      & Native  & \texttt{openat} \\
\texttt{faccessat2} & Syscall & \texttt{openat} \\
\texttt{openat}     & Syscall & \texttt{openat} \\
\texttt{statfs}     & Syscall & \texttt{statfs} \\
\texttt{newfstatat} & Syscall & \texttt{newfstatat} \\
\texttt{newuname}   & Syscall & \texttt{newuname} \\
\texttt{sysinfo}    & Syscall & \texttt{sysinfo} \\
\texttt{execve}     & Syscall & \texttt{execve} \\
\bottomrule
\end{tabularx}
\end{table}

\input{packed/tab-14android}

\input{packed/tab-deeplink}
\input{packed/tab-experimetsetting}

\input{packed/tab-ad-para}

\clearpage
\onecolumn

\input{packed/para-llm-pipedetail}

\end{document}

%% file: section/ethics.tex
\section*{Ethical Considerations}
\label{appendix:ethics}

We follow community best practices to minimize the potential harms of this research.
(1) \textit{Responsible disclosure}. Before submitting this paper, we reported our findings to major companies affected by semi-drive-by splash ads, such as PDD, Kuaishou Technology, ByteDance, Ctrip, Alipay, and Baidu, and communicated the details of our investigation to their security teams.
In response to our report, PDD provided internal confirmation and feedback indicating that our observations align with their internal findings and that they have deployed defensive measures consistent with their business needs.
Kuaishou Technology acknowledged the impact and stated that it remains within an acceptable range for their services.
Alipay confirmed our findings, assigned them a medium-severity rating, and awarded us a bug bounty in recognition of the report.
The other companies are still processing our report internally.
(2) \textit{Minimize the (potential) use of personal data.}
The risk that this study could expose personal data is low, as it collects essentially no personally identifiable information from any real end users.
One potential risk arises from the usage trace synthesis, where the synthesized contacts may contain phone numbers belonging to real users, which may be inadvertently passed to malicious entities (including ad networks).
To minimize this risk, during our large-scale measurement, we generated candidate phone numbers through Company-A's workflow and then used Alibaba Cloud's phone-number status API~\cite{AlibabaCloudCPNS} to select only numbers identified as unassigned before populating them into VMs.
Hence, we believe that the synthesized contacts were not associated with any individuals and would not reveal any real social relationships or communication histories.
(3) \textit{Consideration of the dual-use risks of \tool{}.}
We believe that a complete release could facilitate independent validation and benefit both researchers and industry practitioners, but it could also enable ad networks (or fraudulent apps) engaging in semi-drive-by splash ads to actively develop targeted circumvention strategies against \tool{}, making it more difficult to characterize this new type of threat in the wild.
To reduce this potential risk, we will adopt a controlled-release model.
Specifically, we will create a project webpage that clearly documents the features of \tool{}, provides a terms-of-use agreement, and includes an access request form.
Instead of releasing \tool{} fully to the public, we will provide access only to applicants who verify their identities, are affiliated with academic institutions or companies with established reputations and legitimate needs to combat deceptive advertising practices, articulate a legitimate purpose for using \tool{}, and agree to the terms of use.
We recognize that this controlled-release model cannot completely eliminate the possibility of malicious use of \tool{}, but we consider it a reasonable best-effort measure to reduce this risk.


\section*{Open Science}

To support transparency and reproducibility, we open-source the key tools developed in this work, including the user usage trace simulation tool \texttt{FakeInfo}, the device fingerprint collection tool \texttt{DeviceFP}, and the app \texttt{libc} comparison tools \texttt{Libchecker} and \texttt{compare.py}. The source code is publicly available at:
\url{https://anonymous.4open.science/r/OpensourceforMaliciousAdSDK-5B5E/}.

Due to our collaboration with an industry partner, parts of the deployed system involve proprietary business logic, production integration, and sensitive operational data. We therefore release the research components needed to reproduce the methodology, while excluding partner-specific production components under commercial confidentiality. These exclusions do not affect the reproducibility of the core techniques described in the paper.

\section*{Ethical Usage of LLMs} 

This work involved only limited use of large language models and similar generative AI tools. Specifically, such tools were used for language editing and refinement; for assisting in the generation of random test data, including Chinese surnames and mobile phone numbers, in the FakeInfo module described in section~\ref{method_implementation}; and, in the method implementation described in \ref{method_implementation}, for assisting in determining whether log texts contained suspected advertising attribution parameters and for providing the corresponding rationale. It should be emphasized that all such judgments were manually reviewed and ultimately determined by the author. The generative AI tools were used solely as auxiliary analytical aids and were not employed to perform black-box automatic classification of log data.

%% file: packed/tab-14android.tex
\begin{table*}[t]
\centering
\caption{Survey on Unlocking and Firmware Extraction for Mainstream Brands on Android 14 Phones}
\label{tab:example}

\setlength{\tabcolsep}{4pt}
\begin{tabularx}{\linewidth}{@{}c |c| c| X@{}}
\toprule
\textbf{Brand} & \textbf{Difficulty} & \textbf{Amount} (million) & \textbf{Description} \\
\midrule
HUAWEI~\cite{HuaweiConsumerBG} & 5 & 288 & HarmonyOS prohibits bootloader unlocking, and no public firmware is available. \\
XIAOMI~\cite{XiaomiSmartphoneBrand} & 4 & 121 & Tightened unlocking policy with verification, qualification, quizzes, and a long waiting period. \\
ONEPLUS~\cite{OnePlusBrand} & 3 & 3.2 & Relatively loose unlocking policy, with official unlocking methods provided. \\
OPPO~\cite{opopo_phone_brand} & 4 & 233 & Strict unlocking policy; bootloader unlocking is generally prohibited. \\
\rowcolor{gray!13}
VIVO~\cite{vivoAbout2025} & 3 & 193 & Some models remain unlockable, and after-sales firmware is often independently unpackable.  \\
\bottomrule
\end{tabularx}
\vspace{0.5ex}
\parbox{\linewidth}{\footnotesize
\textit{Note.} The numbers indicate the difficulty of unlocking; the larger the number, the greater the difficulty. Amount refers to the number of users.
}
\end{table*}

%% file: packed/tab-deeplink.tex
\begin{figure*}[!t]
\centering
\small
\caption{Representative ad-related DeepLink patterns.}
\label{tab:specialdeeplinks}
\vspace{0.05in}
\begin{minipage}{0.98\linewidth}
\begin{enumerate}[leftmargin=1.8em,itemsep=5pt,topsep=2pt,parsep=0pt,partopsep=0pt]

\item \textbf{Super-app launch} (\textit{Scheme / Handler:} hap / Quick App; \textit{Structure:} Direct)

{\scriptsize
\dlbase{hap://app/com.dd.library}\dlpath{/pages/spa}\\
\dlparam{?pkg=com.dd.library\&channel\_id=txtth5vo\&link\_id=txtth5vo-ll-ai-h02\&pageExt2=reader\&book\_id=6581\&mediumType=tx\&utm\_term=26532023140}\\
\dlparam{\&utm\_campaign=CREATIVE\_COMPONENTS\_INFO\&utm\_source=DYNAMIC\_CREATIVE\_ID\&utm\_ad\_id=ENCRYPTED\_POSITION\_ID\&qz\_gdt=uuxtc2fsniajuwmysp7q}\\
\dlparam{\&pageDeviceId=174805393452164554\&deviceMod=6\&pageExt1=229ljwsx\&autoPullup=1\&hapType=H5-hap-auto}
}

\item \textbf{Market transit} (\textit{Scheme / Handler:} vmini $\rightarrow$ tbopen $\rightarrow$ https; \textit{Structure:} Nested multi-hop)

{\scriptsize
\dlbase{vmini://vivo.com}\dlpath{/push}
\dlparam{?jumpToTarget=}\\
\dlnested{tbopen://m.taobao.com/tbopen/index.html?h5Url=https://pages-fast.m.taobao.com/wow/z/hdwk/farm-ssr/bargain-backflow?disableNav=YES\&forbidRefineType=goOut}\\
\dlnested{\&shareKey=ddkek1AQRI9MEJ1qnetWX86VVut\_sk=1.aHdS0dXnhB4DANXxGhOpX2s4\_21646297\_1752669605627.CustomQRcode.farmkankanjia}
}

\item \textbf{Market download} (\textit{Scheme / Handler:} market / App Market; \textit{Structure:} Direct)

{\scriptsize
\dlbase{market://details}\dlparam{?id=com.dragon.read}
}

\item \textbf{Market search} (\textit{Scheme / Handler:} vivomarket / Vivo Market; \textit{Structure:} Search redirect)

{\scriptsize
\dlbase{vivomarket://search}\\
\dlparam{?id=\%E5\%B8\%B8\%E8\%AF\%BB\%E5\%85\%8D\%E8\%B4\%B9\%E5\%B0\%8F\%E8\%AF\%B4\&th\_name=ocpc\_appstore\_ttis\&keep\_store=true\&backurl=back\_url\&agent=AUTO}
}

\item \textbf{WeChat jump} (\textit{Scheme / Handler:} weixin / Mini-program; \textit{Structure:} Direct)

{\scriptsize
\dlbase{weixin://dl/business/}\\
\dlparam{?appid=wx888219abee1eea7e\&path=pages/config-page/config-page\&query=id\%3D23946\%26statistical\_from\%3D}
}
\item \textbf{Search redirect} (\textit{Scheme / Handler:} baiduboxapp / Baidu App; \textit{Structure:} Search redirect)

{\scriptsize
\dlbase{baiduboxapp://v1/browser/search}\\
\dlparam{?append=1\&needlog=1\&newwindow=0\&simple=0\&stay=1\&upgrade=1\&query=\%E8\%9E\%BA\%E6\%97\%8B\%E7\%AE\%A1\%E5\%8E\%82\&ad\_id=bddpa\_57591690734136\&ac\_id=52314271\&}

\dlparam{reqid=ehyww6v7zq6vw\&traceid=jjPbNjY7EfCgMCCQb4ujxg\&logargs=\%7B\%22channel\%22\%3A\%221028247s\%22\%2C\%22ext\%22\%3A\%22\%7B\%5C\%22dppage\%5C\%22\%3A\%5C\%22}

\dlparam{search\%5C\%22\%2C\%5C\%22platform\%5C\%22\%3A\%5C\%22feitian\%5C\%22\%2C\%5C\%22sid\%5C\%22\%3A\%5C\%22\%7Bqueryid\%7D\%5C\%22\%7D\%22\%2C\%22from\%22\%3A\%22openbox\%22\%2C\%22}

\dlparam{outerid\%22\%3A\%2257591690734136\%22\%2C\%22page\%22\%3A\%22other\%22\%2C\%22source\%22\%3A\%221028247s\%22\%2C\%22type\%22\%3A\%22\%22\%2C\%22value\%22\%3A\%22url\%22\%7D}
}

\item[*] \textit{URI formatting.} These examples are app-specific DeepLink URIs rather than ordinary web URLs. They use custom schemes resolved by Android app markets, super-app containers, or mini-program runtimes.

\textit{Color notation.} \dlbase{Blue} indicates the base scheme/handler, \dlpath{green} indicates the path component, \dlparam{orange} indicates query parameters and attribution-related fields, and \dlnested{purple} indicates a nested or wrapped redirection target.

\textit{Example.} In Item 2, \dlbase{vmini://vivo.com} denotes the outer handler, \dlpath{/push} denotes the invoked action path, \dlparam{jumpToTarget=} marks the redirection parameter, and the \dlnested{purple segment} denotes the nested downstream target carried inside the outer DeepLink.

\end{enumerate}

\end{minipage}
\end{figure*}

%% file: packed/tab-experimetsetting.tex
\begin{table*}[t]
\centering
\caption{Experimental platforms and configurations.}
\label{tab:platform}

\setlength{\tabcolsep}{4pt}
\begin{tabularx}{\linewidth}{@{}c | c | X@{}}
\toprule
\textbf{Tool} & \textbf{Android Version} & \textbf{Platform Configuration} \\
\midrule

VPBOX & Android 10 & Pixel 3a XL, 4 GB RAM, 64 GB storage. \\

AdHive & Android 14\_r45 & Alibaba Cloud ECS (\texttt{ecs.g8y.16xlarge}, ARM, 64 vCPUs, 256 GiB RAM, 1 TB storage), running Ubuntu 22.04. \\

Redroid\_arm & Android 14\_r45 & Alibaba Cloud ECS (\texttt{ecs.g8yecs-7}, 2 vCPUs, 8 GB RAM, 100 GB storage), running Ubuntu 22.04. \\

Genymotion\_arm & Android 14 & Apple Mac M1 Pro host with 32 GiB RAM and 1 TB storage; Genymotion version 3.9.0. \\

Emulator\_arm & Android 14 & Apple Mac M1 Pro host with 32 GiB RAM and 1 TB storage; Android Studio Panda 2025.3.2. \\

MuMu\_arm & Android 12 & Apple Mac M1 Pro host with 32 GiB RAM and 1 TB storage; MuMu version 1.8.10. \\
Emulator\_x86 & Android 14 & Dell Precision 3480, Intel Core i7-1370P, 64 GB RAM; Android Studio Panda3 2025.3.3 \\
Genymotion\_x86 & Android 14 & Dell Precision 3480, Intel Core i7-1370P, 64 GB RAM; Genymotion version 3.9.0. \\

Real device & Android 14 &  VIVO S17e 
PD2285B\_A\_141.1.170, 12GB RAM, 256 GB storage. \\

\bottomrule
\end{tabularx}
\end{table*}

%% file: packed/tab-ad-para.tex
\begin{table*}[t]
\centering
\caption{Advertising Attribution parameter categories }
\label{tab:ad_feature_categories}

\setlength{\tabcolsep}{4pt}
\begin{tabularx}{\linewidth}{@{}c | X | X@{}}
\toprule
\textbf{Category} & \textbf{Description} & \textbf{Members} \\
\midrule

Source Channel Features & Indicates the media source, channel, or entry point from which the advertising traffic originates & source, src, channel, openFrom, gd\_label, ug\_channel\_source, growth\_channel\_id, spm, bc\_fl\_src, amug\_fl\_src, lch, tra\_from \\

Ad Hierarchy Features & Indicates advertising entities, including accounts, campaigns, ad groups, ad units, creatives, and materials & accountId, account\_id, adgroupId, adId, adid, ad\_id, creativeId, creative\_id, cid, aid, plan\_id, unit\_id, ads\_set, mat\_pkg\_id \\

Click and Request Tracking Features & Used to identify ad clicks, impressions, requests, or conversion chains for attribution and log tracking & click\_id, clickid, reqid, request\_id, requestId, traceid, tracereq\_id, impressionId, pdd\_bid\_id, soid, lpck, exp2\_did \\

Ad Strategy and Billing Features & Represents ad targeting, pricing, bidding strategies, and delivery modes & type, ads\_type, \_p\_ads\_type, typeocpc, chargetype, \_p\_launch\_type, bid\_correct, delay\_bid, bucket\_dsp, eff\_lx \\

\bottomrule
\end{tabularx}
\end{table*}

%% file: packed/para-llm-pipedetail.tex
\section{Pipeline Details}
\label{appendix:raw_log}
\subsection{Tuple description}
To better support automatic data processing, we construct a canonical record for each event, represented as an 8-item tuple: 
\{\texttt{appPkg}, \texttt{callingPkg}, \texttt{action}, \texttt{scheme}, \texttt{host}, \texttt{path}, \texttt{uri}, \texttt{ts}\} . This 8-tuple abstraction is to eliminate formatting differences among different apps, advertising platforms, super apps, and app marketplaces, allowing subsequent expert rules to perform matching and comparison on a unified representation.

In the tuple, \textit{appPkg} denotes the package name explicitly appearing in the DeepLink; \textit{callingPkg} denotes the package name of the app that sends the DeepLink request; \textit{action} indicates whether the event is labeled as \texttt{Normal} or \texttt{Unknown}. \textit{Normal} denotes ordinary app activity, such as opening the homepage or a login page. Because apps also use DeepLinks for normal functionality, not all DeepLink events are suspicious. Here, \textit{Unknown} denotes a state where the event may indicate fraudulent behavior based on preliminary analysis.  \textit{scheme} denotes the URI scheme in the DeepLink; \textit{host} denotes the destination host in the redirected URL; and \textit{path} denotes the intermediate routing path, particularly in \texttt{vmini} cases. The uri denotes the part excluding the scheme. For specific examples, please refer to Appendix~\ref{B:rawlog}. 

\subsection{Raw captured Log}
\label{B:rawlog}
A raw captured log is shown below.\\
{\scriptsize
\texttt{ts: 2025:07:34:45}\\
\texttt{caller: com.app.kge.free}\\
data:
\dlbase{ksnebula://search}
\dlparam{?keyword=\%E5\%90\%83\%E7\%88\%86\%E8\%BE\%A3\%E7\%BE\%8E\%E9\%A3\%9F\&source=}\\
\dlparam{EXT\_mix\&openFrom=ANDROID\_GDT\_TX\_NRXXLZT\_CPC\_DIY46780\&type=laxin\&inner\_cid=472990428}\\
\dlparam{\&creativeId=472990428\&impressionId=dd2fig4nefvdm01\&accountId=57320057\&adgroupId=36253253675}\\
\dlparam{\&adId=\_\_DYNAMIC\_CREATIVE\_ID\_\_}
}

 In this example, \textit{ksnebula://} denotes the DeepLink scheme, and \textit{caller} corresponds to \textit{callingPkg}. Since no explicit destination host or intermediate routing path is present in this log, the corresponding \textit{host} and \textit{path} fields are set to null. In addition, because the DeepLink does not explicitly contain an application package name, \textit{appPkg} is also set to null.

The scheme \textit{ksnebula} is associated with the \textit{Kuaishou} application and is used to launch the app. However, the DeepLink request is issued by \textit{com.app.kge.free}, which attempts to invoke another application through the \textit{ksnebula} scheme. This cross-application invocation may indicate potentially fraudulent behavior under the preliminary analysis criteria. Therefore, the \textit{action} field is labeled as \textit{unknown}. The corresponding tuple is shown below.

{\scriptsize
\noindent
\begin{tabularx}{\columnwidth}{@{}lX@{}}
\texttt{appPkg}     & \texttt{null} \\
\texttt{callingPkg} & \texttt{com.app.kge.free} \\
\texttt{action}     & \texttt{unknown} \\
\texttt{scheme}     & \texttt{ksnebula} \\
\texttt{host}       & \texttt{null} \\
\texttt{path}       & \texttt{null} \\
\texttt{uri}      & \texttt{search?keyword=\%E5\%90\%83\%E7\%88\%86\%E8\%BE\%A3\%E7\%BE\%8E\%E9\%A3\%9F\&source=EXT\_mix\&} \\
                    & \texttt{openFrom=ANDROID\_GDT\_TX\_NRXXLZT\_CPC\_DIY46780\&type=laxin\&} \\
                    & \texttt{inner\_cid=472990428\&ug\_channel\_source=UG\_CSXXL\&} \\
                    & \texttt{growth\_channel\_id=UG\_CSXXL\&creativeId=472990428\&} \\
                    & \texttt{impressionId=dd2fig4nefvdm01\&accountId=57320057\&} \\
                    & \texttt{adgroupId=36253253675\&adId=\_\_DYNAMIC\_CREATIVE\_ID\_\_} \\
\texttt{ts}         & \texttt{2025:07:34:45}
\end{tabularx}
}

\subsection{Tag details}
\label{tagrule}
Based on our analysis of 100 canonical record tuples, we identify five important features. The features are grouped into a 5-item tuple: \texttt{(cross\_app, scheme\_mismatch\_appPkg, has\_attribution, is\_search, is\_jump)} to describe suspicious advertising-related DeepLink behavior from structural, semantic, and statistical views. The specific rules defined by experts are presented as follows.

The \textit{cross\_app} field represents whether a DeepLink is initiated by one app but points to another app. We set it to \textit{True} when \textit{callingPkg} does not match the app family indicated by the DeepLink scheme. This field does not prove ad fraud by itself, but it captures a necessary condition for many ad redirections, user acquisition, re-engagement, and attribution flows. In contrast, normal in-app navigation is usually initiated by the target app itself and uses its own scheme.

The \textit{scheme\_mismatch\_appPkg} field represents the inconsistency between the scheme and the package name in the parameters. Normal DeepLinks usually follow the form \texttt{targetapp\_scheme://path?parameter}, and their parameters do not contain an extra package name. However, DeepLinks involving \texttt{Super-app Launch}, \texttt{Market Download} or \texttt{WeChat Jump} often use a container scheme, while the actual target is encoded in the parameters, host, or path. This is aimed at these special event types, such as \textit{App Launch}, \textit{Super-app Launch}, \textit{Market Transit}, \textit{Market Download}, \textit{Market Search}, or \textit{WeChat Jump}. The specific event types are explained in Figure~\ref{tab:specialdeeplinks}. Therefore, we set \textit{scheme\_mismatch\_appPkg} to \textit{True} when \textit{appPkg} does not match the scheme, or when \textit{appPkg} is null and the DeepLink exhibits characteristics of one of these \textit{special} event types.

The \textit{has\_attribution} field represents whether the uri contains advertising attribution parameters, such as \textit{source}, \textit{openFrom}, \textit{creativeId}, \textit{impressionId}, \textit{accountId}, \textit{adgroupId}, \textit{adId}, \textit{click\_id}, or \textit{traceid}. These fields identify ad channels, campaign objects, creatives, clicks, requests, impressions, or billing strategies, so their presence suggests that the DeepLink is part of an ad measurement or attribution flow. As shown in Table~\ref{tab:ad_feature_categories}, we summarized common advertising attribution parameters into several categories from these 100 logs.

The \textit{is\_search} field represents whether this DeepLink is used to open another app for search.  The criterion is whether the DeepLink contains any of the following patterns.

\url{oaps://mk/search?kw=}

\url{vivomarket://search?id=}

\url{baiduboxapp://v1/browser/search}

The \textit{is\_jump} field  represents whether DeepLink involves an intermediate redirect or jump. We set it to True when the DeepLink contains the following pattern: \url{push ?jumpToTarget=}.
Since advertising attribution parameters are diverse and often customized by different advertising SDKs, a purely rule-based approach may miss non-standard parameter expressions. To improve the coverage of attribution-parameter identification, we introduce GPT-4o as an auxiliary semantic checker.

Specifically, when the expert rules classify \textit{has\_attribution}, \textit{is\_search} and \textit{is\_jump} as \textit{False}, \tool{} sends the \textit{uri} field in the 8-tuple, together with a fixed prompt, to GPT-4o. The goal is to identify potential attribution-related parameters that are not covered by the existing expert rules. GPT-4o is required to return a structured response with a Boolean decision and the concrete parameters supporting its judgment. All positive suggestions are manually reviewed before being added to the expert rulebook. Therefore, GPT-4o is used only to assist rule refinement, rather than serving as a standalone detector.

The prompt used for this semantic checking step is as follows:

\begin{quote}
You are an expert in semantic analysis of DeepLink URIs. I will provide the DeepLink URI. Please analyze whether it contains parameters related to advertising attribution.

First, fully decode the uri component, including any nested encoded content.

Second, semantically inspect all decoded keys and values, and determine whether any of them are related to advertising attribution, such as campaign identifiers, creative identifiers, impression identifiers, click identifiers, account identifiers, ad group identifiers, channel sources, or billing-related fields.

If attribution-related parameters are found, return a JSON object in the following format:

\texttt{\{"state": true, "reason": "..."\}}

Otherwise return a JSON object in the following format:
\texttt{\{"state": false, "reason": "..."\}}

The \texttt{reason} field should specify the concrete parameters that support your judgment, so that the result can be manually reviewed.
\end{quote}

\subsection{Rule-based Evidence Aggregation}
\label{evidence}

We define the following aggregation rules.
\textbf{R1} indicates that if a log satisfies the cross-application feature and carries advertising attribution-related parameters, it can be classified as a high-confidence advertising log.
\textbf{R2} indicates that if a log satisfies the cross-application feature and contains a search-related pattern, it can be classified as a high-confidence advertising search log.
\textbf{R3} indicates that if a log satisfies the cross-application feature, and the package name or App ID found in the log can be used to determine that it belongs to one of the following categories: Super-app, Market Download, or WeChat Jump, then the log can be classified accordingly with high confidence. \textbf{R4} indicates that if a log does not satisfy the cross-application feature but contains advertising attribution parameters, it suggests that the app has opened its own advertisement. Such a log can be classified as a high-confidence app self-launch advertising log. \textbf{R5} indicates that if a log only contains non-cross-application features, it can be classified as a high-confidence non-advertising log. \textbf{R6} indicates that the log satisfies the cross-application feature and that the associated DeepLink contains a jump or redirection. It can be classified as a high-confidence Market transit log. 

The union of the results identified by R1, R2, R3 and R6 represents the high-confidence fraudulent logs extracted from a mixture of normal logs and fraudulent logs. This interpretation is specific to our honeypot setting: the device performs no user interaction or manual click, so a cross-application advertisement-related DeepLink automatically triggered by an app cannot be attributed to normal user behavior. Therefore, under our threat model, detecting the existence of such a log is sufficient to identify high-confidence fraudulent behavior.

\begin{align*}
R_1:\quad & \mathsf{cross\_app} \land \mathsf{has\_attribution}
\rightarrow \text{advertisement-related launch} \\
R_2:\quad & \mathsf{cross\_app} \land \mathsf{is\_search}
\rightarrow \text{market-search }\\
R_3:\quad & \mathsf{cross\_app} \land \mathsf{scheme\_mismatch\_appPkg}
\rightarrow \text{Super-app
Launch / Market Download / WeChat Jump} \\
R_4:\quad & \neg \mathsf{cross\_app} \land \mathsf{has\_attribution}
\rightarrow \text{self-launched advertisement event}.\\
R_5:\quad & \neg \textit{cross\_app} \land \neg \textit{has\_attribution} \rightarrow \neg \text{high-confidence fraudulent log}\\
R_6:\quad & \textit{cross\_app} \land \textit{is\_jump} \rightarrow \text{Market-transit}
\end{align*}

We evaluate whether the proposed pipeline analyzer can reliably identify high-confidence advertising logs. The evaluation is designed around three goals: validating the correctness of the conservative aggregation rules, measuring the contribution of each feature, and assessing the benefit of GPT-4o-assisted rule expansion.

\subsection{Details of Pipeline Evaluation}
\label{piplinedetails}

Two annotators independently labeled the same 200 logs using the original DeepLink URI, the decoded URI string, and the normalized 8-field record. The annotation categories include benign log, advertisement-related launch, market-search launch, special transit launch and self-launched advertisement event. After labeling, disagreements were resolved through discussion. The annotators achieved an initial raw agreement of 92\%, and the resolved labels were used as ground truth. We then compared the labels generated by our rule-based pipeline against the manually verified labels. We reported precision, recall, F1 score, and false positive rate.
\subsection{Prediction Distribution}

\label{appendix:prediction}

To further understand where the improvement comes from, we analyze the category-level prediction distribution. However, our evidence aggregation rules are not mutually exclusive: a single event may satisfy multiple rules. For example, a market-search DeepLink may also carry advertisement attribution parameters and therefore trigger both the generic attribution rule R1 and the market-search rule R2. Directly counting all matched rules would over-count events and make the category distribution misleading. 

To avoid duplicate counting, we separate evidence matching from final label assignment. During evidence matching, the analyzer records all rules triggered by each event. For reporting and evaluation, each event is assigned exactly one primary label using a deterministic specificity-first priority order. The intuition is that R1 captures a generic cross-application advertisement-related launch, while R2, R6, and R3 capture more specific behavioral subtypes, including market-search, market-transit, and special transit launches. Therefore, when multiple rules are triggered, the more specific behavior label is preferred over the generic advertisement-related label. For cross-application events, we use the priority order R2 > R6 > R3 > R1. Thus, if an event satisfies both R1 and R2, it is reported as a market-search event rather than a generic advertisement-related launch. If an event satisfies both R6 and R3, it is reported as Market transit because explicit jump or redirection semantics provide a more specific characterization than a general scheme-application mismatch. This priority order is used only to produce mutually exclusive category-level statistics; it does not change the underlying evidence matching. The total number of high-confidence fraudulent logs is computed over unique events satisfying any of R1, R2, R3, or R6. 

Table~\ref{tab:gpt_rule_expansion_distribution} reports the resulting mutually exclusive prediction distribution before and after GPT-4o-assisted rule expansion. As shown in Table~\ref{tab:gpt_rule_expansion_distribution}, GPT-4o-assisted rule expansion mainly reduces unclassified fraudulent logs from 21 to 9 by converting them into advertisement-related launches, while leaving benign-log predictions unchanged.

\begin{table}[t]
\centering
\caption{Prediction distribution before and after GPT-4o-assisted rule expansion.}
\label{tab:gpt_rule_expansion_distribution}
\begin{tabular}{lcc}
\toprule
Predicted label & Expert rules only & Expert rules + GPT-4o \\
\midrule
\multicolumn{3}{l}{\textbf{Fraudulent logs} $(n=100)$} \\
Advertisement-related launch & 65 & 77 \\
Super-app launch & 7 & 7 \\
Market-search & 6 & 6 \\
Market-transit & 1 & 1 \\
Unclassified & 21 & 9 \\
\midrule
High-confidence fraudulent logs & 79 & 91 \\
\midrule
\multicolumn{3}{l}{\textbf{Benign logs} $(n=100)$} \\
Non-advertisement launch & 95 & 95 \\
Self-launched advertisement event & 5 & 5 \\
High-confidence fraudulent logs & 0 & 0 \\
\bottomrule
\end{tabular}
\end{table}
\subsection{Ablation evaluation}
\label{appendix:ablation}
To understand the contribution of each feature, we conduct an ablation study by removing one feature at a time. Specifically, we evaluate five variants, each removing one of \textit{cross\_app}, \textit{scheme\_mismatch\_appPkg}, \textit{has\_attribution}, \textit{is\_search}, and \textit{is\_jump}. This experiment quantifies the contribution of each feature to the final labeling performance.

Table~\ref{tab:pipeline_ablation}  shows that \texttt{cross\_app} and \texttt{has\_attribution} are the key features. Removing cross\_app reduces recall from 91.0\% to 0.0\%, while removing has\_attribution reduces it to 14.0\%. Other features mainly improve category-level interpretation, since disabling \texttt{is\_search}, \texttt{scheme\_mismatch\_appPkg}, or \texttt{is\_jump} either causes only a small recall drop or collapses specific categories into the generic advertisement-related label. The false-positive rate remains 0.0\% across all evaluated variants, supporting the conservative design of our analyzer.

\begin{table*}[t]
\centering
\caption{Ablation study for high-confidence fraudulent log detection.}
\label{tab:pipeline_ablation}
\begin{tabular}{lccccccccc}
\toprule
Variant & TP & FP & FN & TN & Precision & Recall & F1 & FPR & Accuracy \\
\midrule
Full pipeline & 91 & 0 & 9 & 100 & 100.0\% & 91.0\% & 95.29\% & 0.0\% & 95.5\% \\
w/o \texttt{cross\_app} & 0 & 0 & 100 & 100 & N/A & 0.0\% & 0.0\% & 0.0\% & 50.0\% \\
w/o \texttt{has\_attribution} & 14 & 0 & 86 & 100 & 100.0\% & 14.0\% & 24.56\% & 0.0\% & 57.0\% \\
w/o \texttt{is\_search} & 89 & 0 & 11 & 100 & 100.0\% & 89.0\% & 94.18\% & 0.0\% & 94.5\% \\
w/o \texttt{scheme\_mismatch\_appPkg} & 91 & 0 & 9 & 100 & 100.0\% & 91.0\% & 95.29\% & 0.0\% & 95.5\% \\
w/o \texttt{is\_jump} & 91 & 0 & 9 & 100 & 100.0\% & 91.0\% & 95.29\% & 0.0\% & 95.5\% \\
\bottomrule
\end{tabular}

\vspace{0.5ex}
\parbox{0.95\textwidth}{\footnotesize
\textit{Note.} Precision is marked as N/A for w/o \texttt{cross\_app} because no event is predicted as a high-confidence fraudulent log. Self-launched advertisement events are not counted as high-confidence fraudulent logs because they do not satisfy the cross-application condition in our threat model. 
}
\end{table*}